\documentclass[a4paper,fleqn,usenatbib,useAMS]{mnras}
\usepackage{graphicx}	
\usepackage{amsmath}	
\usepackage{amssymb}	
\usepackage{multicol}        
\usepackage{bm}		
\usepackage{pdflscape}	
\usepackage[T1]{fontenc}
\usepackage{ae,aecompl}
\usepackage{color}
\title {{The large scale structure formation in an expanding Universe}}
 \author[Hameeda et al.]{\small{Mir Hameeda$^{1,2}$\thanks{E-mail: hme123eda@gmail.com}, Behnam Pourhassan$^{3}$\thanks{E-mail: b.pourhassan@du.ac.ir},
 Syed Masood$^{4}$\thanks{E-mail: masood@zju.edu.cn}, Mir Faizal$^{5,6,7}$\thanks{E-mail: mirfaizalmir@googlemail.com},
 Li-Gang Wang$^{4}$\thanks{E-mail: sxwlg@yahoo.com}, Shohaib Abass$^{8}$\thanks{E-mail: shohaibabass@gmail.com}}\\
$^{1}$Department of Physics, S.P. Collage, Srinagar 190001, Kashmir,  India\\
$^{2}$Inter University Centre for Astronomy and Astrophysics, Pune, India\\
$^{3}$School of Physics, Damghan University, Damghan, 3671641167, Iran\\
$^{4}$Zhejiang Province Key Laboratory of Quantum Technology and Device, Department of Physics, Zhejiang University, Hangzhou 310027, China\\
$^{5}$Irving K. Barber School of Arts and Sciences, University of British Columbia, Kelowna, British Columbia, V1V 1V7, Canada\\
$^{6}$Department of Physics and Astronomy, University of Lethbridge, Lethbridge, Alberta, T1K 3M4, Canada\\
$^{7}$Canadian Quantum Research Center, 204-3002 32 Ave Vernon, BC V1T 2L7 Canada\\
$^{8}$Department of Physics, Central University of Kashmir, Tulmulla Campus Ganderbal 191131, Kashmir, India }

\begin{document}
\maketitle

\begin{abstract}
In this paper, we analyze the effects of expansion on large scale structure formation in our Universe. We do that by incorporating a cosmological constant term  in the gravitational partition function.  This gravitational partition function with a cosmological constant is used for analyzing the thermodynamics of this system. We analyze the virial expansion for this system, and obtain its equation of state. It is observed that the generalization of this equation of state
is like the Van der Waals equation. We also analyze a gravitational phase transition in this system using  the mean field theory. We construct the cosmic energy equation for this system of galaxies, and discuss its consequences.  We  obtain and analyze the distribution function for this system, using the gravitational partition function. We also  compare the results obtained in this paper with the observational data.
\end{abstract}

\begingroup
\let\clearpage\relax
\endgroup
\newpage

\section{Introduction}

The clustering of galaxies  is responsible for forming  large scale structure in our Universe   (\citealt{PEEBLES}; \citealt{PEEBLESb}; \citealt{clus}). So, it is very important to analyze the clustering of galaxies, and this can be done using  simulation  (\citealt{a1}; \citealt{a2}), or by relating various physical parameters to observational data (\citealt{a3}; \citealt{a4}). In these works, the local matter distribution in galaxies or clusters is analyzed (either by simulation or using observation). However, for analyzing the formation of large scale structure in the Universe, it is more appropriate to approximate each galaxy as a point, analogous to a point particle in a statistical mechanical system. This approximation is valid as the size of galaxies is much smaller than the distance between them.
Thus, it is possible to write a statistical mechanical gravitational partition function for this system, and use it for analyzing  the large scale structure formation  (\citealt{sas}; \citealt{sas2}).
It may be noted that this formalism has been used to investigate the thermodynamic description of the cosmological many-body problem and  galaxy clustering (\citealt{cosa1}; \citealt{cosa2}). Here the thermodynamic quantities, like temperature, are obtained using the kinetic theory of gases, with each galaxy acting as a particle analog.
It has been observed that  such a system can be analyzed using a  quasi-equilibrium description, as    the    macroscopic quantities  change slowly compared to local relaxation time scales  (\citealt{cos1}; \citealt{cos2}; \citealt{cos3}; \citealt{cos4}).
As the extended structure of a galaxy is approximated by a point particle,  this gravitational partition function can diverge. However, these divergences  can be removed by using a  softening parameter  (\citealt{ahm10}).  This softening parameter incorporates the effects of having an extended structure in the gravitational partition function.
This   softening parameter also modifies the  thermodynamic fluctuations in this system. So, the correct thermodynamics for this system  has to be analyzed  using  this gravitational partition function modified by the softening parameter (\citealt{sas84}).
The clustering of a system of galaxies has  been studied using the  grand canonical ensemble of galaxies  (\citealt{ahm02}).  This has been done by analyzing the   distribution functions  and moments of distributions, such as their skewness and kurtosis.  The study of such  distribution functions for  galaxies has indicated that galaxy clusters are surrounded by individual  halos (\citealt{10}; \citealt{20}).
The  gravitational partition function has also been used to obtain the   specific heats and  isothermal compressibility for such a system (\citealt{ahm06}).

It may be noted that the clustering occurs due to gravitational interaction between different galaxies. However, as the gravity pulls the galaxies towards each other (causing clustering), the  expansion of the Universe is expected to move the galaxies away from each other. It is now known that the Universe is undergoing accelerated expansion. This is based on observations of Type Ia Supernovae (SNeIa) (\citealt{1ab}; \citealt{6ab}). It may be noted that even though the effects from the cosmological constant  are usually   neglected on   astrophysical scales, it is possible for the  cosmological constant to have an impact  at the  scale of galaxy clusters  (\citealt{cosd1}; \citealt{cosd2}; \citealt{cosd3}; \citealt{cosd4}; \citealt{cosd5}). Such effects  have been studied  in models of interacting  dark matter and dark energy (\citealt{cosd1}; \citealt{cosd2}). It has been observed in such interacting models that the dark energy can be constrained from equilibrium states of local galaxy clusters (\citealt{cosd51}).
In fact, in such models, it has been argued that local effects from cosmological constant at the scale of galaxy clusters can have important astrophysical consequences (\citealt{cosd6}). It has been demonstrated that cosmological constant can change the  dynamics of a  group of galaxies  (\citealt{cosd71}). This analysis has been done using  five  groups of galaxies and  the Virgo cluster.
It has been argued that cosmological constant can play an important role in the formation of galaxy clusters (\citealt{cosd7}).
The  effect of dark energy on  local galaxy clusters has been studied using the Fisher matrix formalism (\citealt{cosd8}). So, the local galaxy clusters contain information about the cosmological constant. In fact, it has been argued that  the value of the cosmological constant can be constrained from an accurate measurement of  the   statistical properties of galaxy clusters  (\citealt{cosd9}).

Thus, it is important to analyze the effects of the cosmological constant on the statistical mechanical approach to the clustering of galaxies. The effect of the cosmological constant on the statistical distribution function of  galaxies has been studied, and this analysis has been used to constrain the value of the cosmological constant (\citealt{cosd91}).
As such a statistical distribution function can be obtained from the gravitational partition function, the  effects of   a cosmological constant term have been also incorporated  in the gravitational partition function (\citealt{1b}).
This gravitational partition function has  been used to obtain the  Helmholtz free energy for a system of galaxies. This Helmholtz free energy  has in turn been used to obtain the entropy of this system. The thermodynamics of this system is used to obtain the
  clustering parameter for this system, and is used for analyzing the effects of the cosmological constant on the clustering of galaxies. As the internal energy of this system depends on the clustering parameter, which in turn depends on the cosmological constant, the dependence of the internal energy on the cosmological constant has also been studied for this system. Finally,  the distribution function for a system of galaxies in an expanding Universe   is obtained using the  grand canonical  gravitational partition.
As it is possible to analyze cosmological models with a dynamical time-dependent  cosmological  constant  (\citealt{cc22, cc44}), it is important to analyze the effects of such a time-dependent cosmological constant  on the clustering of galaxies. Thus, the  gravitational partition function with a time-dependent cosmological constant  has also been constructed, and it has been used to study the effects of dynamical dark energy on structure formation in our Universe   (\citealt{1}).  For this model with  dynamical dark energy, Helmholtz free energy  has been  used to obtain the entropy of the system, which in turn has been  used to obtain the dependence of the clustering parameter on the dynamical dark energy. The correlation function between galaxies has also been calculated and found to be consistent with observations.

It is possible to modify general relativity by adding  higher powers of the curvature tensor to obtain  $f(R)$ gravity, and this modification of general relativity also modifies the large-distance behavior of the gravitational potential (\citealt{gravity}).
Such a model of  $f(R)$ gravity has also been  used to study the expansion of the Universe (\citealt{gravity}). The gravitational partition function for $f(R)$ gravity has been used to analyze the effects of $f(R)$ gravity on the large scale structure formation  (\citealt{2}). It was observed that this modification of the gravitational partition function is consistent with observations. The thermodynamics of such a system of galaxies interacting through the modified  gravitational potential  of $f(R)$ gravity, has also been studied. It has also been demonstrated that $f(R)$ gravity can be constrained using  the PLANCK data on galaxy clusters (\citealt{2az}). This was done by
  calculating the  pressure profiles of different  galaxy clusters. It was assumed that this gas of galaxies was in hydrostatic equilibrium within the  $f(R)$ gravitational potential well. It was observed that the profile of this system of galaxies   fits the observation  data, without requiring dark matter.
The thermodynamics of a system of galaxies has also been analyzed using MOND,
 and this was done by analyzing the modifications to the gravitational partition function from MOND (\citealt{3}).   It was observed that the  modification of the gravitational partition function from MOND also modified  the thermodynamic potential for this system, which in turn modified the formation of large scale structure. The modification to the gravitational partition function from  MOG
has also been studied, and it was observed that the clustering of galaxies depends on the large scale modifications to Newtonian potential in MOG (\citealt{4}). It has been done by analyzing the thermodynamics  of  a system of galaxies interacting through MOG Newtonian potential (\citealt{4}). The clustering in brane-world models has also been studied using a modification to gravitational partition function (\citealt{Hameeda2}). This was done  by analyzing the modification to Newtonian potential from super-light brane world perturbative modes. This modified Newtonian potential modifies the thermodynamics of the system, and this 
changes the large scale structure formation. Thus, it was possible to analyze the effects of super-light brane world perturbative modes on the large scale structure formation in our Universe.

It may be noted that it is also possible to study the gravitational phase transition for a system of galaxies using gravitational partition function (\citealt{sin12}; \citealt{sin14}; \citealt{2b}). It was observed that a first order phase transition occurs when this system of galaxies starts to cluster from an initial homogeneous phase to a phase with a large scale structure. The  phase transition in a  system of galaxies has also been analyzed using the   gravitational  partition function as a function of complex fugacity (\citealt{4b}). This was done by extending  the Yang-Lee theory to the gravitational phase transition. It was observed that masses of individual galaxies can affect the formation of large scale structure.
As it is important to consider the effects of  the cosmic expansion  on the structure formation, we will analyze the gravitational  phase transition for a system of galaxies, with a cosmological constant term.
We would like to point  out that it is possible to study the clustering in a system of galaxies using
cosmic energy equation   (\citealt{Ahmad}). Even though the cosmic energy equation is obtained by assuming galaxies as point particles (\citealt{Ahmad}),  the modification to the cosmic energy equation from the extended structure of galaxies has also been studied (\citealt{ce}). The cosmic energy equation can be used for analyzing the dependence of clustering on the gravitational potential.    In fact, the effects of a large distance modification to  gravitational  potential  on clustering have also been analyzed   using a modified cosmic energy equation  (\citealt{Hameeda}). As it is important to use the  gravitational partition function modified by  the cosmological constant, in this paper, we will also analyze the cosmic energy equation using such a modified gravitational partition function.

\section{Gravitational Partition Function}
It is known that  galaxies  interacting via gravitation are   not in equilibrium. However, it is possible to analyze such a cosmological system using  quasi-equilibrium (\citealt{cos1}; \citealt{cos2}; \citealt{cos3}; \citealt{cos4}). This is because for  such a cosmological system, the   macroscopic quantities  change slowly compared to local relaxation timescales, and hence they can be studied using quasi-equilibrium relations. These macroscopic quantities  include the average temperature, which is obtained by using the kinetic theory of gases, with each galaxy represented by a particle analog.  The pressure and density are also  such  macroscopic quantities.  Now the average density of the Universe     is  $ 3.5 \times  10^{10} m_\odot Mpc^{-3}$, whereas  most galaxies  have an average mass of about
  $10^{11}m_\odot$ (\citealt{cos41}).  The   dynamical timescale of  the Universe   is $ \tau = 25
Gyr$. There are about   $0.35$ galaxies per cubic megaparsec, with an  average peculiar velocity of
   $Mpc Gyr^{-1}$.  So, the time for a galaxy to cross a cell of   $3000$ galaxies would be around  $20 Gyr$ (\citealt{cos1}). However, local  time scales for such a system will be much shorter. The density of a  cluster of galaxies, like our local group of galaxies, is about  $1.5 \times 10^{12} m_\odot Mpc^{-3}$. This is several  times greater than the density of the Universe, and so the  local group has a dynamical timescale $\tau =  4 Gyr $.  As most galaxy clusters have a  diameter of $ 2-4 Mpc$, the time for a galaxy to cross them is about  $3 Gyr$ (\citealt{cos1}). Thus, the microscopic perturbations on local scales will relax  much  faster than the macroscopic  scale (\citealt{cos2}). So, this system can evolve from one   equilibrium state to another, and can be analyzed using such a quasi-equilibrium description (\citealt{cos3}; \citealt{cos4}).

So, we can analyze this system, using a large number of galaxies  distributed  in an ensemble of cells, all of them with same volume $V$, or radius $R_1$ and average density $\rho$.   Both the number of galaxies and their total energies vary among these cells and hence it can be appropriately represented by a grand canonical ensemble.
These galaxies within the system have pairwise gravitational  interactions generated by the modified Newtonian potential. It is  further assumed that the distribution is statistically homogeneous over large regions. Now the temperature for such a system can  be  obtained from the average kinetic energy of galaxies, using the kinetic theory of gases, with each  galaxy representing a particle analog of such a gas. We can thus obtain  an average temperature $T$ from a kinetic theory of galaxies, and use it to construct the gravitational partition function.  So, the  gravitational partition function of a system of $N$ galaxies  of mass $m$ interacting through the modified gravitational potential  energy $\Phi$ can be written as (\citealt{1}; \citealt{2}; \citealt{3}; \citealt{4}),
\begin{eqnarray}
&&Z(T,V)= \frac{1}{\Lambda_1^{3N}N!}\int d^{3N}p\hspace{2mm}d^{3N}r\nonumber \\
&\times&  \exp\left(-\frac{\left[\sum_{i=1}^{N}\frac{p_{i}^2}{2m}\right]+\Phi(r_{1}, r_{2}, r_{3},\dots, r_{N}, t)}{T}\right),
\end{eqnarray}
where $\Lambda_1$ is the mean thermal wavelength which is defined as $(2\pi m T)^{-\frac{1}{2}}$, and  $p_{i}$ is the momentum for different galaxies.
Integrating momentum space integral for the phase space, we obtain the following expression
\begin{eqnarray}
Z_N(T,V)=\frac{1}{N!}\left(\frac{1}{\Lambda_{1}^{2}}\right)^{3N/2}Q_N(T,V).
\end{eqnarray}
Here  $Q_{N}(T,V)$ is the configurational integral of the system, and  is  given by
\begin{eqnarray}
Q_{N}(T,V)=\int....\int \prod_{1\le i<j\le N} \exp[-\frac{\Phi_{ij}}{T}]d^{3N}r.
\end{eqnarray}
The gravitational potential energy $\Phi_{ij}\equiv\Phi(r_{1}, r_{2}, \dots, r_{N},t)$ is a function of the relative position vector $r_{ij}=|r_{i}-r_{j}|$ (because the two-body force is a central one), and the cosmological time (because of its dependence on the cosmological expansion $a(t)$). The total potential energy is the sum of the potential energies of all pairs.

This potential energy $\Phi(r_{1}, r_{2}, \dots, r_{N} ,t)$ can be expressed as
\begin{eqnarray}
\Phi(r_{1}, r_{2}, \dots, r_{N},t)&=&\sum_{1\le i<j\le N}\Phi(r_{ij},t)\nonumber\\
&=&\sum_{1\le i<j\le N}\Phi(r,t).
\end{eqnarray}
We can use a two-particle Mayer function
$f_{ij}=e^{-(\Phi_{ij})/T}-1$ to analyze the non-ideal case. This function vanishes in absence of interactions (ideal case), and is non-zero for interacting galaxies. Thus, the configurational integral can be written as
\begin{eqnarray}
Q_{N}(T,V)&=& \int....\int (1+f_{12})(1+f_{13})(1+f_{23})(1+f_{14})\nonumber\\
&&\dots (1+f_{N-1,N}) d^{3}r_{1}\hspace{2mm}d^{3}r_{2}\dots d^{3}r_{N}.
\end{eqnarray}

It is known that the cosmological constant can  change the statistical properties of the system of galaxies  (\citealt{cosd9}), such  as  the statistical distribution function  (\citealt{cosd91}).
As the  statistical distribution function can be obtained from the gravitational partition function, it is important to incorporate    cosmological constant term  in the gravitational partition function.
So,  we modify the above gravitational potential energy by introducing a cosmological constant term $\Lambda$ along with an extra parameter  that depends on the change in the scale factor $a$.
Thus, we  write the  modified potential energy for this system of galaxies  as (\citealt{2010})
\begin{eqnarray}\label{potential}
\Phi(r_{ij},t)=-\frac{Gm^2}{r_{ij}}-\frac{m\Lambda r_{ij}^2}{6}+\frac{m{\ddot{a}}r_{ij}^2}{2a},
\end{eqnarray}
where  $G$ is Newton's gravitational constant.
Moreover, due to the extended nature of galaxies, the softening parameter $\epsilon$ (\citealt{ahm10}; \citealt{sas84}; \citealt{ahm02}; \citealt{10}; \citealt{20}; \citealt{ahm06}) is incorporated  in this potential energy term as
\begin{eqnarray}\label{pot}
\Phi(r_{ij},t)=-\frac{Gm^2}{(r_{ij}^2+\epsilon^2)^{1/2}}-\frac{m\Lambda r_{ij}^2}{6}+\frac{m\ddot{a}r_{ij}^2}{2a}.
\end{eqnarray}
Now the  two-particle Mayer function can be obtained from this   expression for the modified potential term. Here, using an approximation for  the weak interaction, the effects from the last term for the Mayer function can  be  neglected. Thus,  by  retaining only the first term,   the   two-particle Mayer function   modified by the cosmological constant term,  can be expressed as
\begin{eqnarray}
f_{ij}=\left(\frac{Gm^2}{T(r_{ij}^2+\epsilon^2)^{1/2}}+\frac{m\Lambda r_{ij}^2}{6T}-\frac{m\ddot{a}r_{ij}^2}{2aT}\right).
\end{eqnarray}

\begin{figure}
 \begin{center}$
 \begin{array}{cccc}
\includegraphics[width=60 mm]{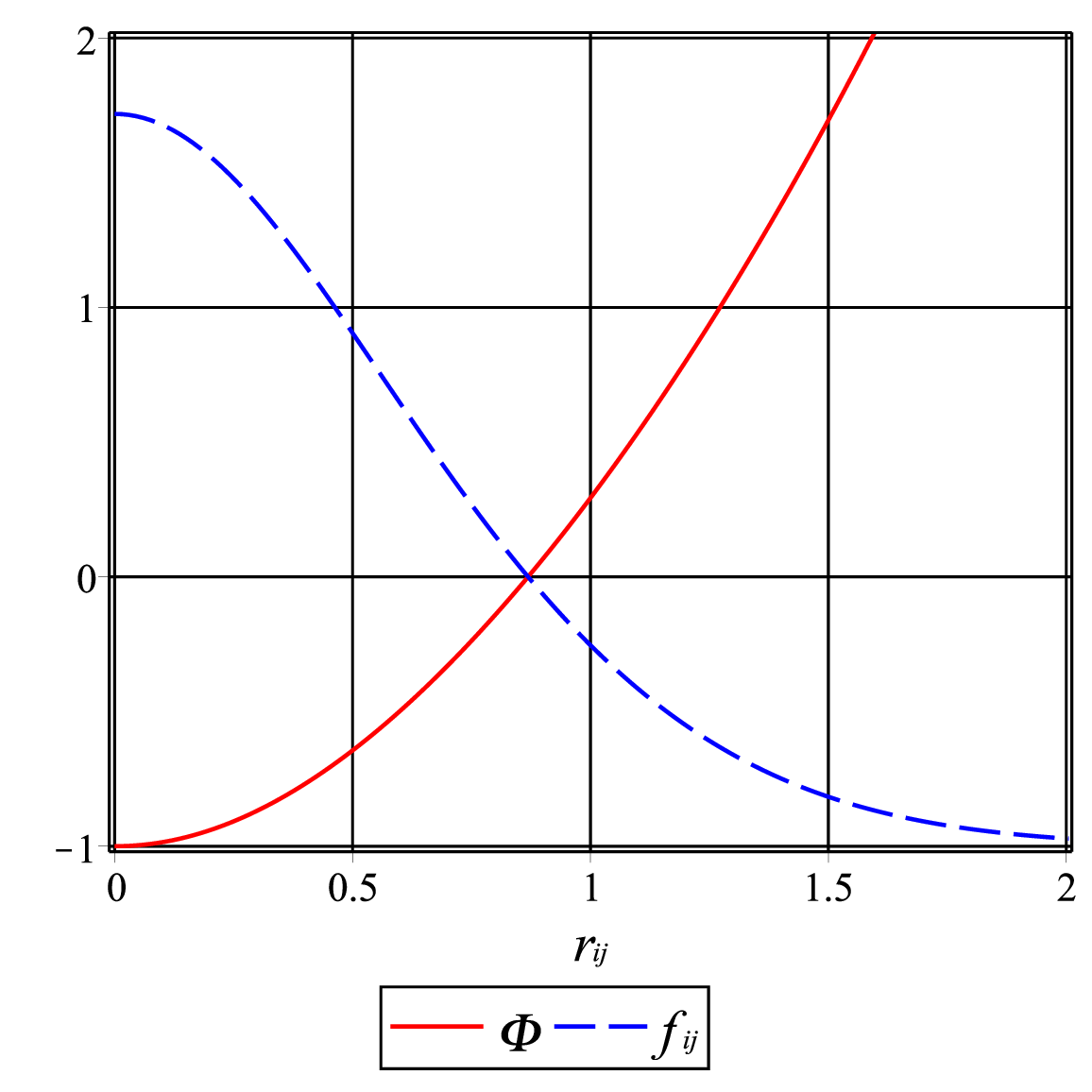}
 \end{array}$
 \end{center}
\caption{ Behavior of the potential energy (solid red line) and the corresponding Mayer function (dashed blue line) versus $r_{ij}$ for $m=G=T=\epsilon=1$ and $\frac{\ddot{a}}{2a}-\frac{\Lambda}{6}=1$. }
 \label{fig1}
\end{figure}

In Fig. \ref{fig1}, we can see the behavior of potential energy (\ref{pot}) and Mayer function in terms of $r_{ij}$. We observe from  the blue dashed line that $f_{ij}$ is bounded in all regions, which produces minus one at infinity and 
has finite positive values at the origin. It is observed that both $\Phi$ and $f_{ij}$ are zero at a single point, as was physically expected.

Now we can  analyze  the case with $N=2$. For this case, we can write the $Q_{2}(T,V)$ as
\begin{eqnarray}
Q_{2}(T,V)&=& 4\pi V\left(\int_{0}^{R_{1}}r^2dr\right)\nonumber\\
&+& 4\pi V\left(\frac{Gm^2}{T}\int_{0}^{R_{1}}\frac{r^2dr}{(r^{2}+\epsilon^{2})^{1/2}}\right)\nonumber\\
&+& 4\pi V\left(\left(\frac{\Lambda m}{6T}-\frac{m\ddot{a}}{2a}\right)\int_{0}^{R_{1}}r^4dr\right).
\end{eqnarray}
Solving the above integral, we obtain the following expression for $Q_{2}(T,V)$
\begin{eqnarray}
Q_{2}(T,V)=V^{2}(1+\alpha x),
\end{eqnarray}
where we defined
\begin{eqnarray}\label{alpha}
\alpha&=&\sqrt{1+\frac{\epsilon^{2}}{R_{1}^{2}}}+\frac{\epsilon^{2}}{R_{1}^{2}}\ln{\left(\frac{\epsilon}{R_{1}+\sqrt{R_{1}^{2}+\epsilon^{2}}}\right)}
\nonumber \\ && +\frac{2R_{1}^{3}}{5Gm}\left(\frac{\Lambda}{6}-\frac{\ddot{a}}{2a}\right).
\end{eqnarray}
Here $\alpha$ depends on the free parameters, such as the softening parameter and radius of the cell $R_1$. We also observe that a   term in $\alpha$ depends explicitly on the value of the cosmological constant. So,  the effect of cosmic expansion on the clustering of galaxies is incorporated  through $\alpha$. We will observe that $\alpha$ plays an important role in clustering, and as it depends explicitly on the cosmological constant, the clustering will also depend on the rate of  cosmic expansion, as is  physically expected.
Here $x$ is obtained by first noting that
$
{3Gm^{2}}/{2R_{1}T}={3Gm^{2}}/{2\rho^{-1/3}T}.
$
Then by using the scale invariance,  $\rho\to\lambda^{-3}\rho$, $T\to\lambda^{-1}T$ and $r\to\lambda r$, we obtain the expression for  $x$ as
\begin{eqnarray}\label{x000}
x = \frac{3}{2}(Gm^2)^{3}\rho T^{-3}=\beta\rho T^{-3},
\end{eqnarray}
where $\beta= \frac{3}{2}(Gm^{2})^{3}$.
Thus, the  configurational integral is a function of the average  temperature   due to the dependence of $x$ on $T$. This expression for $x$ can be used to obtain an explicit expression for any configurational integral. It may be noted that even
 though such an expression  has been derived for $Q_{2}(T,V)$, the same procedure can be used to evaluate any other configurational integral.

\begin{figure}
 \begin{center}$
 \begin{array}{cccc}
\includegraphics[width=60 mm]{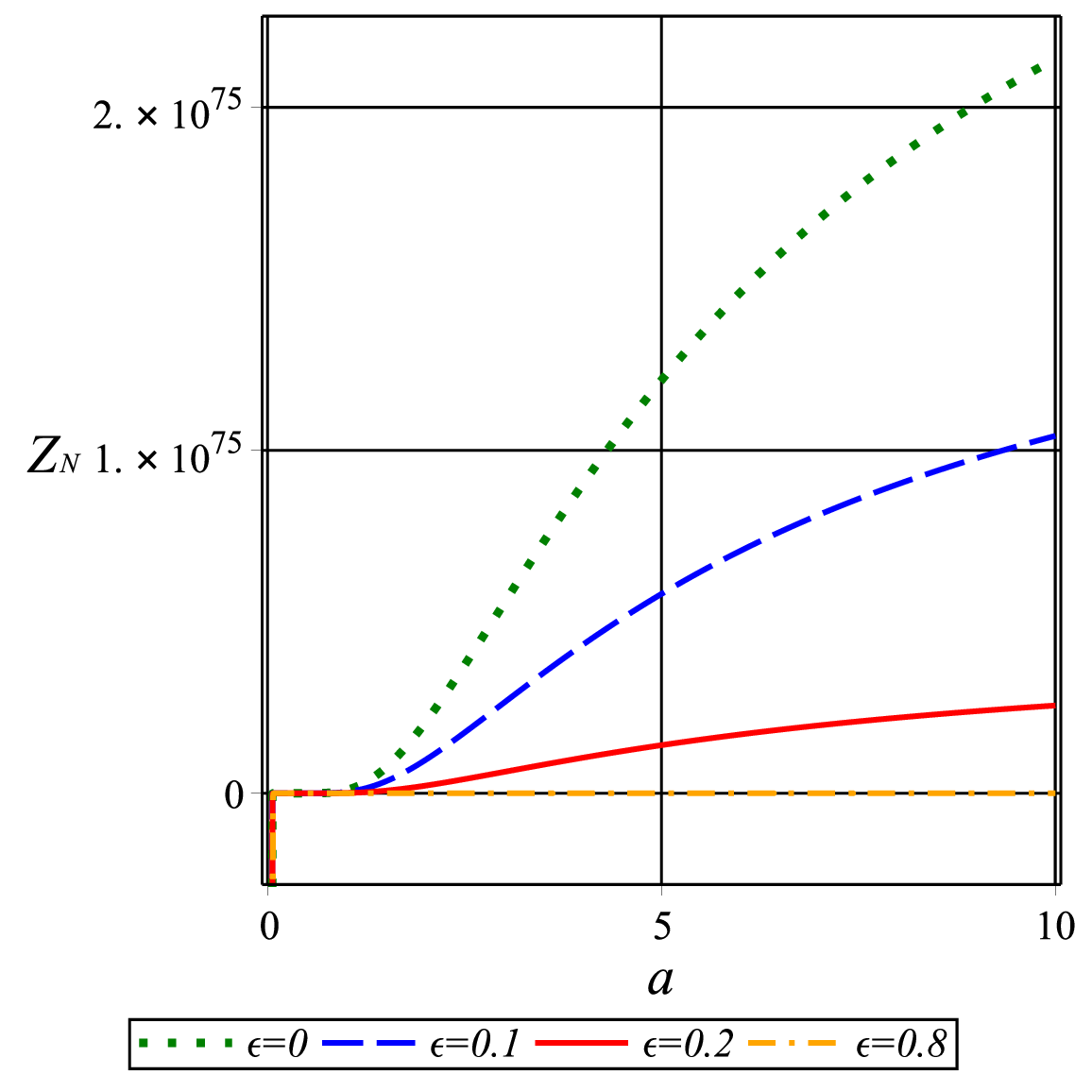}
 \end{array}$
 \end{center}
\caption{ Behavior of the partition function per unit volume versus $a$, for $m=G=T=R_{1}=\Lambda=\ddot{a}=1$ and $N=50$. }
 \label{figz}
\end{figure}

Thus, by following the same procedure, we can write  the  general configurational integral as
\begin{eqnarray}
Q_{N}(T,V)=V^{N}(1+\alpha x)^{N-1}.
\end{eqnarray}
Here we again observe that the  general configurational integral  also depends on the value of the cosmological constant, due to the dependence of $\alpha$ on the cosmological constant.
Using this general configurational integral, we can write  the gravitational partition function for a system of galaxies  as
\begin{eqnarray}\label{ZN111}
Z_N(T,V)=\frac{1}{N!}(2\pi m T)^{3N/2}V^{N}(1+\alpha x)^{N-1}.
\end{eqnarray}
This gravitational  partition function depends on the cosmological constant, due to the dependence of   $\alpha$  on the cosmological constant. We can use  Fig. \ref{figz} to investigate  the  effects of the scale factor on the gravitational partition function. We note that by increasing  the value of the scale factor, the  gravitational  partition function also increases. We would like to point out that the clustering of galaxies can be studied using the  gravitational partition function for a system in  quasi-equilibrium  (\citealt{cos1}; \citealt{cos2}; \citealt{cos3}; \citealt{cos4}). Now as we have  obtained the modification to the gravitational partition function from  the cosmological constant term,   we can  analyze the effect of the cosmological constant on  modified thermodynamics using this modified gravitational partition function. This is  valid, as the    macroscopic quantities would still change slowly compared with local relaxation timescales, and so this  system can again  be analyzed using a quasi-equilibrium state  (\citealt{1b}).

\section{Thermodynamics}
The gravitational partition function obtained  in the previous section can be used to  analyze   the  thermodynamic behavior of this system of interacting galaxies in an expanding Universe. We can also analyze the  dependence of such a system on the scale factor  for   power law cosmology. We first  write the internal energy of this system of galaxies  as
\begin{equation}\label{U0}
U=T^{2}\frac{d\ln{Z_{N}}}{dT},
\end{equation}
where $Z_{N}$ is given by equation (\ref{ZN111}). Using the Eq.  (\ref{U0}), we obtain  the following expression (assuming $m=G=1$),
\begin{equation}\label{U1}
U=\frac{3T\left[3\ddot{a}R_{1}^{5}-aU_{N}\right]}
{3\ddot{a}R_{1}^{5}-aU_d},
\end{equation}
where  $U_N$ and $U_d$ are given by
\begin{eqnarray}\label{U2}
U_N&=&10NR_{1}^{5}T^{3}+\Lambda R_{1}^{5}+15R_{1}\sqrt{R_{1}^{2}+\epsilon^{2}}\nonumber\\
&+&15\epsilon^{2}\ln{\left(\frac{\epsilon}{R_{1}+\sqrt{R_{1}^{2}+\epsilon^{2}}}\right)},\nonumber\\
U_d&=&10R_{1}^{5}T^{3}+\Lambda R_{1}^{5}+15R_{1}\sqrt{R_{1}^{2}+\epsilon^{2}}\nonumber\\
&+&15\epsilon^{2}\ln{\left(\frac{\epsilon}{R_{1}+\sqrt{R_{1}^{2}+\epsilon^{2}}}\right)}.
\end{eqnarray}
It may be noted that the internal energy of this system depends on the softening parameter, which has been introduced to incorporate the extended nature of the galaxies. This is expected as the internal energy of a thermodynamic system of extended structures is expected to be different from the internal energy of a similar thermodynamic system of point-like structures. The internal energy of the system also depends on the cosmological constant, which is again expected, as we expect the internal energy of such a system of galaxies to depend on the rate of cosmic expansion.

\begin{figure}
 \begin{center}$
 \begin{array}{cccc}
\includegraphics[width=65 mm]{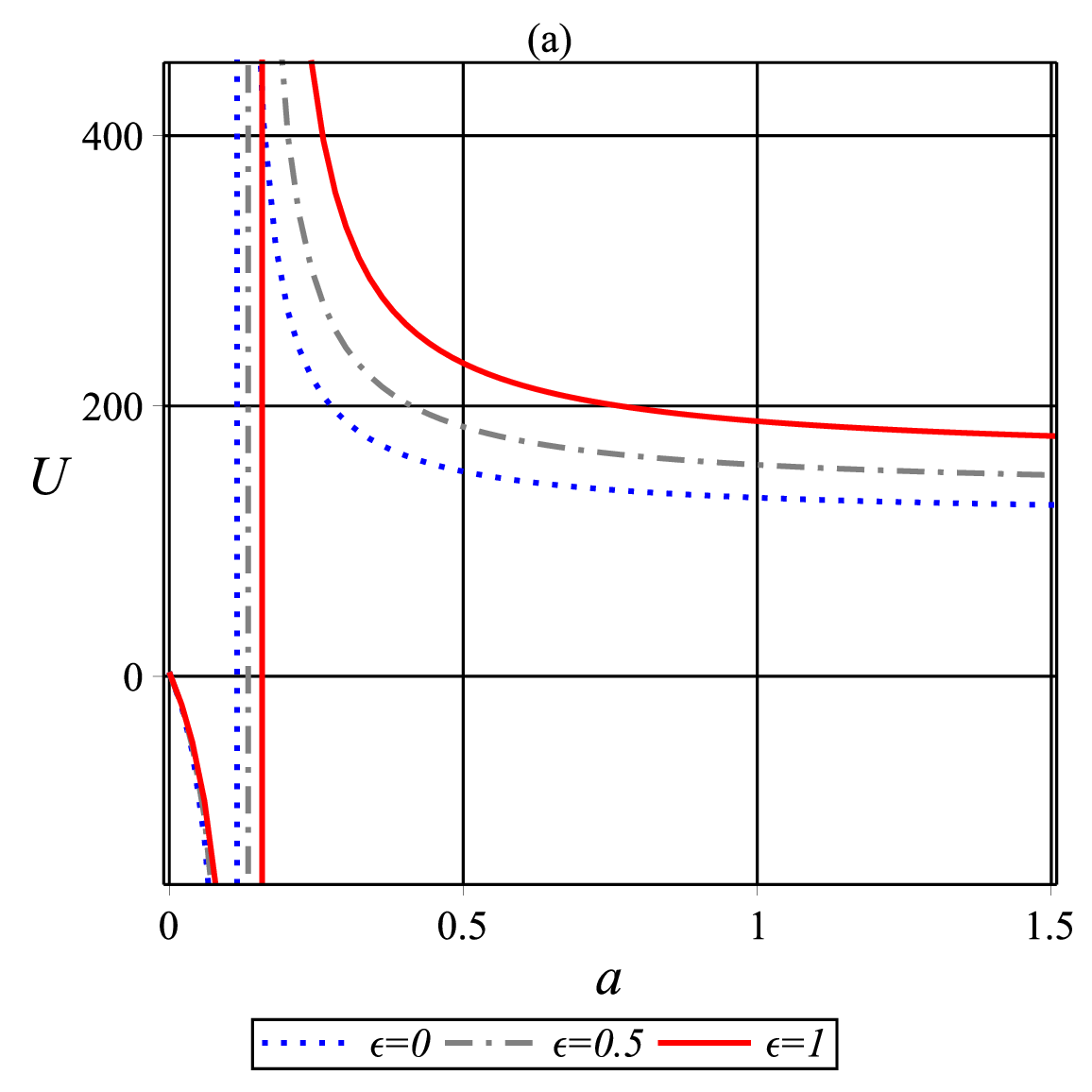}\\
\includegraphics[width=65 mm]{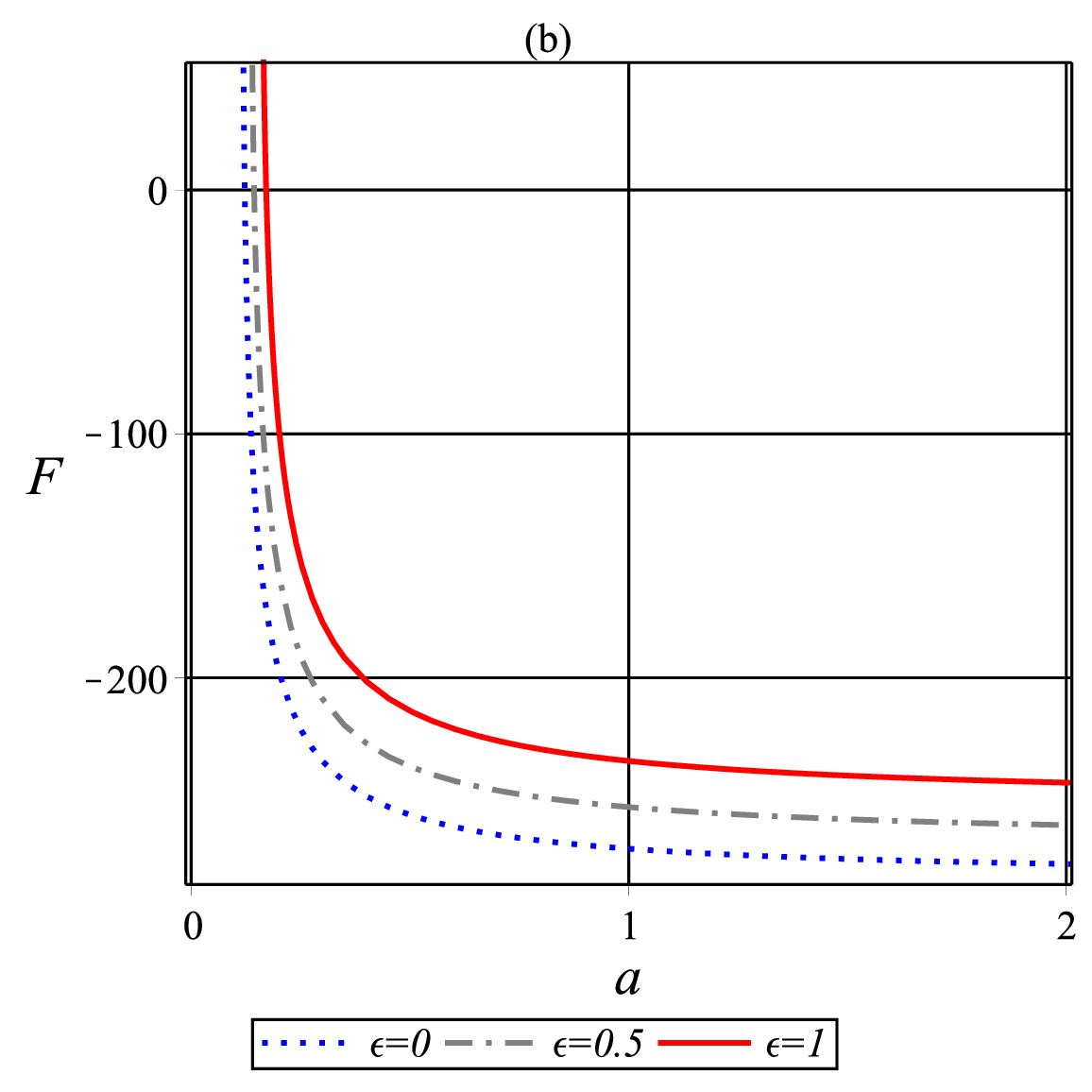}
 \end{array}$
 \end{center}
\caption{Behavior of the (a) internal energy and (b) Helmholtz free energy per unit volume versus $a$ for $m=G=T=R_{1}=\Lambda=\ddot{a}=1$ and $N=100$.}
 \label{figE}
\end{figure}

We can analyze the behavior of the Helmholtz free energy using the gravitational partition function modified by the cosmological constant term, $Z_{N}$. Thus, using $Z_{N}$ from Eq. (\ref{ZN111}), we can write the Helmholtz free energy modified by the cosmological constant term as
\begin{equation}\label{F0}
F=-T\ln{Z_{N}}.
\end{equation}
We observe that for $N=1$, the internal energy is related to temperature as  $U=3T$.  So,  the  internal energy of this system resembles the  internal energy of three dimensional harmonic oscillator (in units of Boltzmann constant), and this behavior is  expected from the  equipartition theorem. In Fig. \ref{figE} (a), we  analyze  the dependence of the internal energy $U$ on the  scale factor $a$. We can observe that there is  a singular point for a small value of the  scale factor. The internal energy is initially negative, and  then becomes  positive, indicating  a phase transition. This phase transition corresponds to the maximum value of Helmholtz free energy (see Fig. \ref{figE} (b)), which  diverges  at the initial stage.
Thus, it seems from the thermodynamic behavior of this system, that there is a phase transition  in it. We would like to point out that it has been suggested that clustering  can be   regarded as a   phase transition (\citealt{sin12}; \citealt{sin14}). Here we observe that to be the case from the behavior of its internal energy and Helmholtz free energy.
We can also observe  from Fig. \ref{figE} (b),  the Helmholtz free energy becomes minimum at  late times.  This indicates that the  resulting configuration is stable. Thus,  with the increasing scale factor, galaxy clusters tend to stabilize. This is expected as the galaxies are expected to cluster, with the expansion of the Universe.

The negative value of the internal energy at the small scale factor or singular point at the  initial stage is related to the distance between   galaxies. It indicates  that there is a minimum value for $R_{1}$ where the system will be stable. We can understand this better by analyzing the entropy $S$  of the system.  Entropy $S$ for a  system of galaxies in an expanding Universe   can be obtained as
\begin{equation}\label{S0}
S=\frac{U}{T}+\ln{Z_{N}}.
\end{equation}

\begin{figure}
 \begin{center}$
 \begin{array}{cccc}
\includegraphics[width=65 mm]{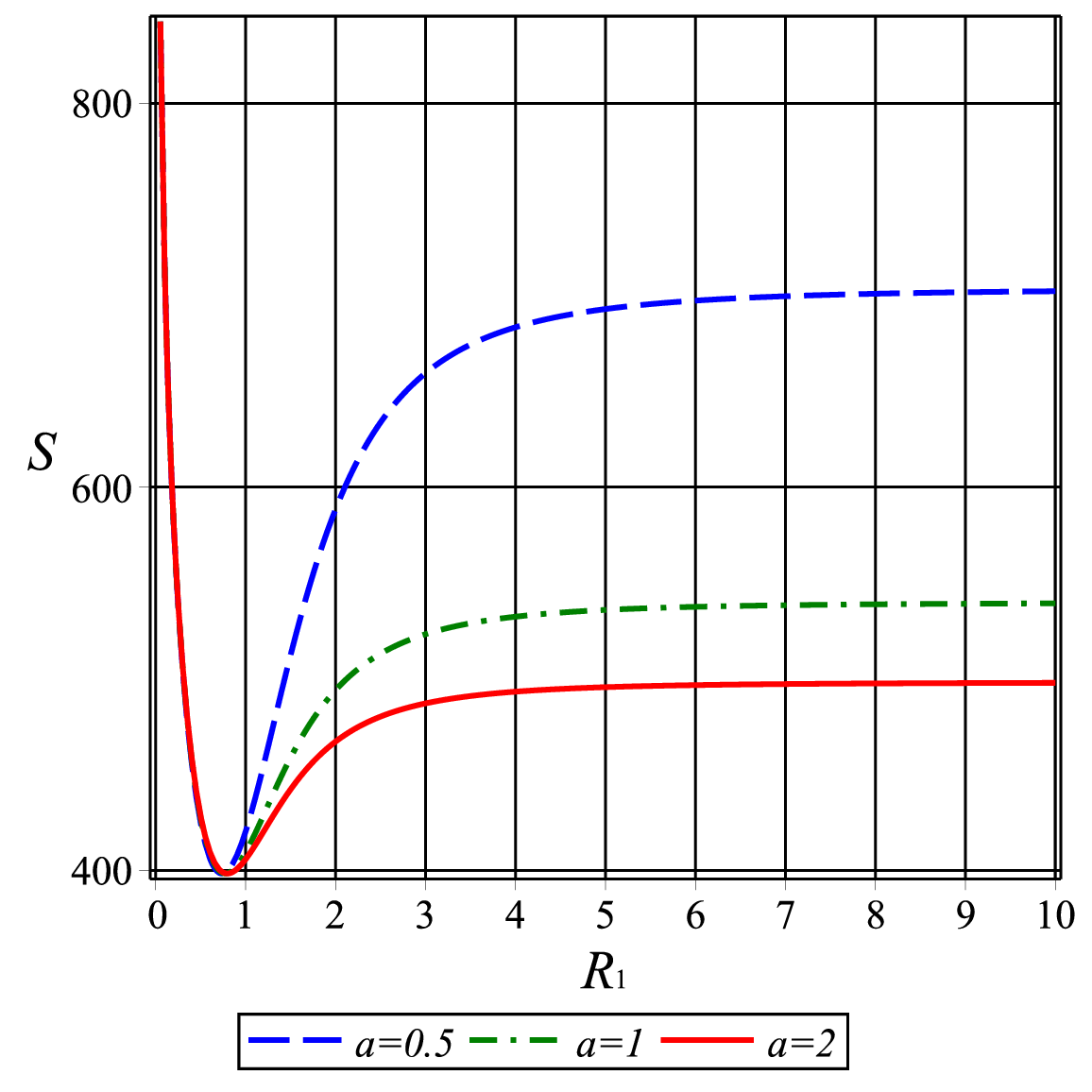}
 \end{array}$
 \end{center}
\caption{Behavior of the entropy per unit volume versus $R_{1}$ for $m=G=T=\Lambda=\ddot{a}=1$, $\epsilon=0.5$ and $N=100$.}
 \label{figS}
\end{figure}

\begin{figure}
 \begin{center}$
 \begin{array}{cccc}
\includegraphics[width=65 mm]{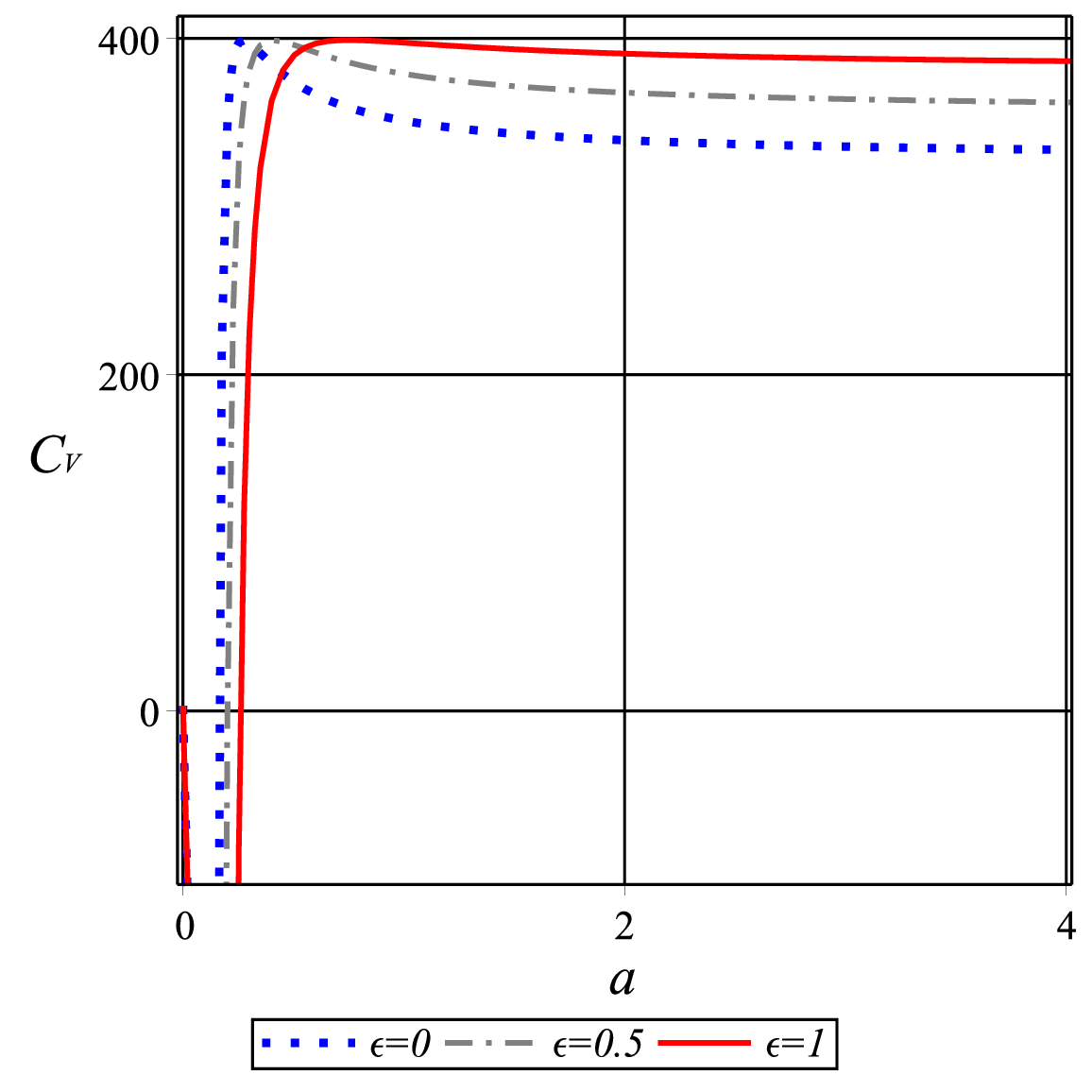}
 \end{array}$
 \end{center}
\caption{Behavior of the specific heat per unit volume versus $a$ for $m=G=T=R_{1}=\Lambda=\ddot{a}=1$ and $N=100$.}
 \label{figC}
\end{figure}

Using  Fig. \ref{figS}, we can analyze the  behavior of the  entropy in terms of $R_{1}$, for various values of the scale factor $a$. We observe that the  entropy is positive, indicating that the resulting configuration is  physical. However, for the smaller values of $R_{1}$,  entropy is a decreasing function of time. We obtain  a critical value of $R_{1}$ as $R_{c}\approx0.8$, and for any  $R_{1}\geq R_{c}$,  the entropy is an increasing function of time,  indicating that the approximation is valid for $R_{1}\geq R_{c}$. Thus, it seems that this approximation is only valid up to a critical size of the cell, and below such a value, the physical system cannot be expressed by this approximation.  A  stable phase for the system is attained due to   the accelerating expansion of the Universe. This can also be seen by analyzing the   heat capacity of this system. The heat capacity  at constant volume $C_{V}$ of this system can be expressed as
\begin{equation}\label{C0}
C_{V}=\left(\frac{dU}{dT}\right)_{V}.
\end{equation}
The behavior  of the specific heat at constant volume is  given by Fig. \ref{figC}.  We can observe  that the specific heat is negative at the early stages.  Also, we can observe  that the heat capacity rises to a maximum, which is like a Schottky anomaly (appears in some of the two-level systems).

\begin{figure}
 \begin{center}$
 \begin{array}{cccc}
\includegraphics[width=65 mm]{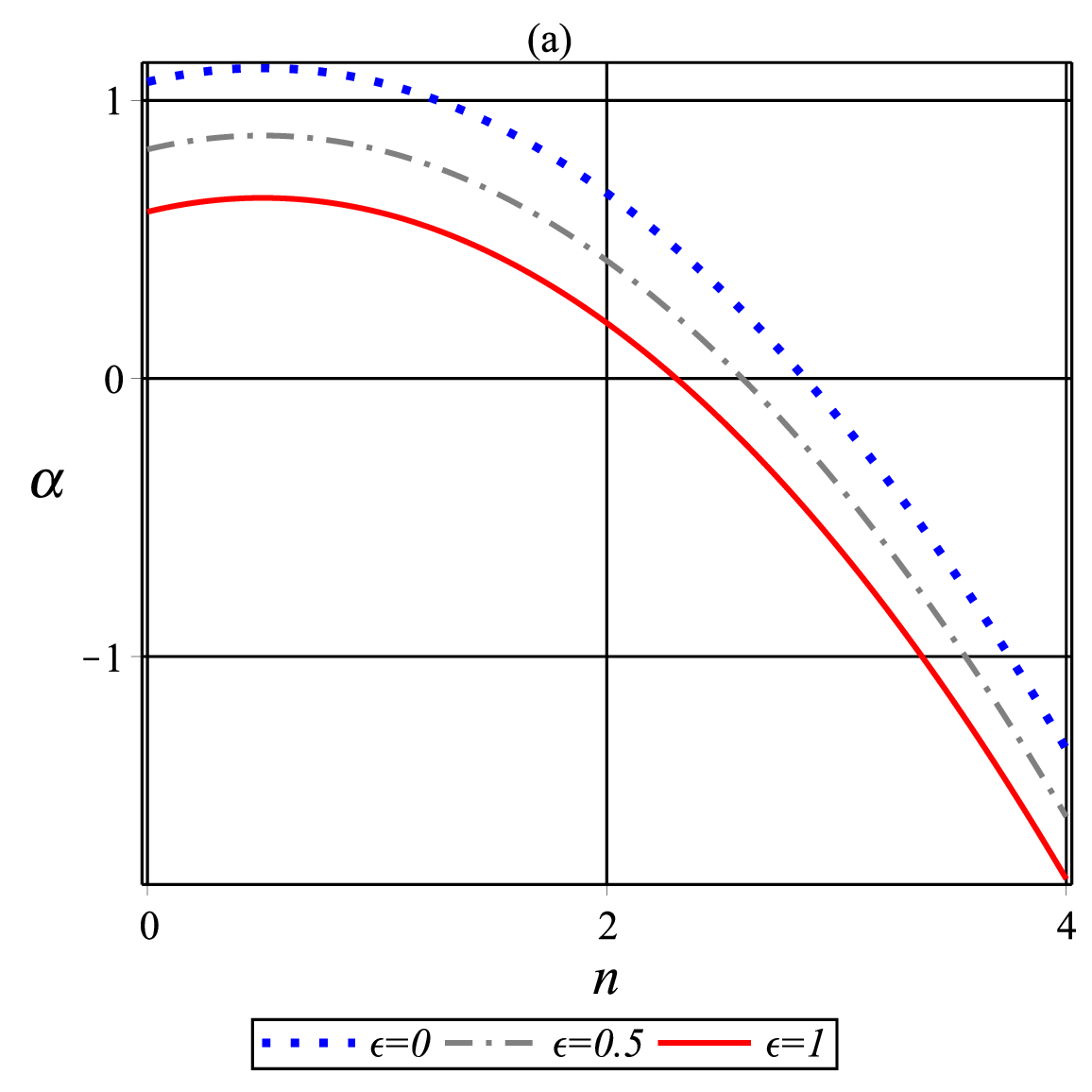}\\
\includegraphics[width=65 mm]{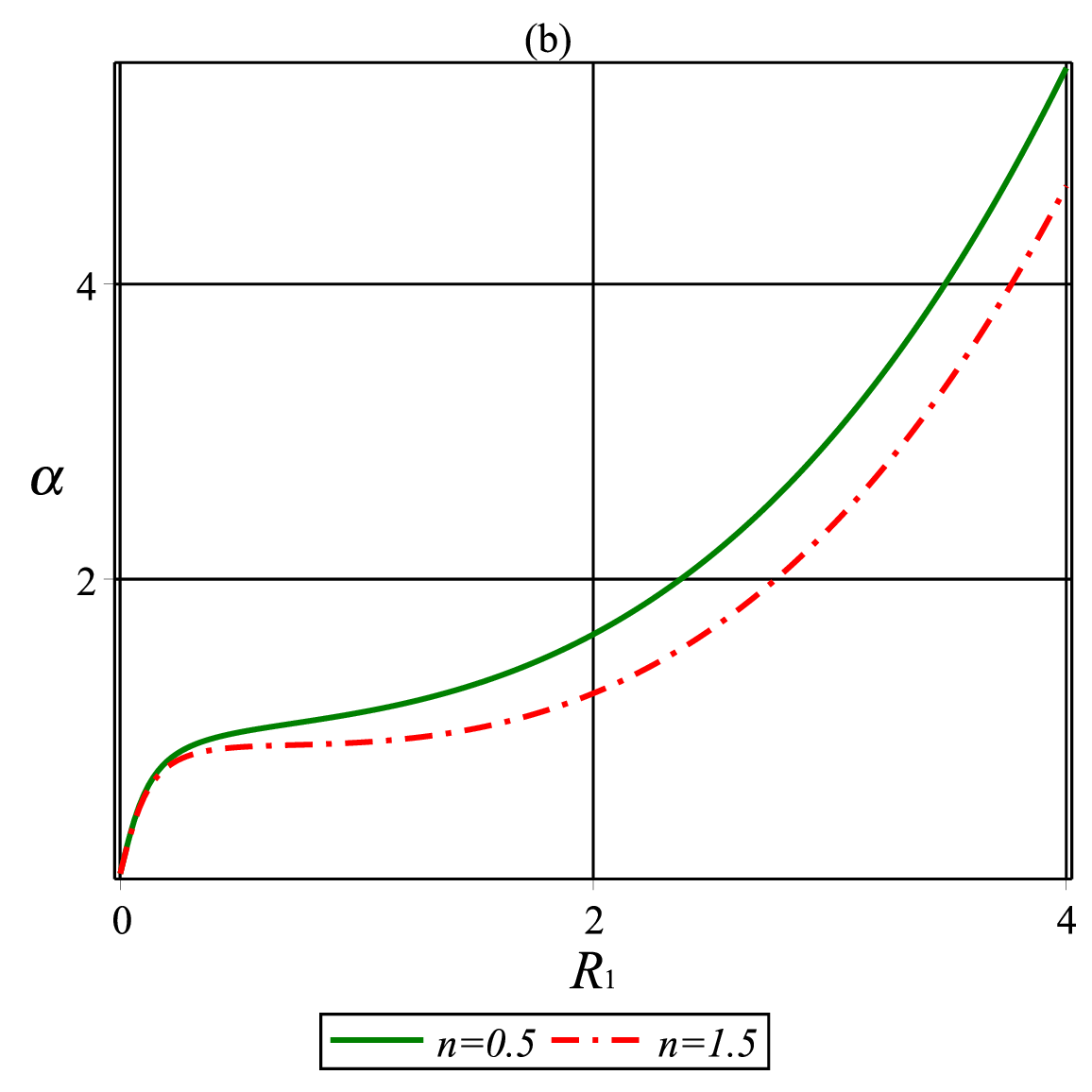}
 \end{array}$
 \end{center}
\caption{ Behavior of $\alpha$ for $m=G=\Lambda=1$. (a) In terms of $n$ for $R_{1}=1$;  (b) In terms of $R_{1}$ for $\epsilon=0.1$.}
 \label{fig7}
\end{figure}

Now we discuss  the special cases, by choosing specific forms of time-dependence for the scale factor.
We can consider the special case of power law dependence of the scale factor,
\begin{equation}\label{scale}
a=a_{0}t^{n},
\end{equation}
where $a_{0}$ is a constant and $n$ is a real number. In a Friedmann-Lema\^{i}tre-Robertson-Walker Universe, the scale factor $a$ gives us the value of $n$,  so that in the radiation-dominated Universe, $n=\frac{1}{2}$, while in a matter-dominated Universe, $n=\frac{2}{3}$.  In those cases, one can obtain Hubble expansion parameter as $H\propto\frac{1}{t}$. On the other hand,  holographic dark energy models suggest that Hubble parameter $H$ is proportional to the length, $H\propto\frac{1}{R_{1}}$. Under this assumption, equation (\ref{scale}) changes to
\begin{equation}\label{scale1}
a=a_{1}R_{1}^{n},
\end{equation}
where $a_{1}$ is another constant. Hence, we can obtain,
$
\ddot{a}=n(n-1){R_{1}^{-2}}a.
$
Now for this case, the Eq.  (\ref{alpha}) can be expressed as
\begin{eqnarray}\label{alpha1}
\alpha&=&\sqrt{1+\frac{\epsilon^{2}}{R_{1}^{2}}}+\frac{\epsilon^{2}}{R_{1}^{2}}\ln{\left(\frac{\epsilon}{R_{1}+\sqrt{R_{1}^{2}+\epsilon^{2}}}\right)}\nonumber\\
&+&\frac{2R_{1}^{3}}{5}\left(\frac{\Lambda}{6}-\frac{n(n-1)}{2R_{1}^{2}}\right),
\end{eqnarray}
where we assumed $G=m=1$. As the gravitational  partition function is proportional to $\alpha$, it is important to analyze the behavior of $\alpha$ in such models.  This is because $\alpha$ depends on the cosmological constant, and this cosmological constant can be used to analyze  the effects of the dark matter and dark energy, in models of interacting dark matter and dark energy  (\citealt{cosd1}; \citealt{cosd2}; \citealt{cosd3}; \citealt{cosd4}; \citealt{cosd5}).  In the plots of Fig. \ref{fig7}, we observe the behavior of $\alpha$ for various values of $n$ and $R_{1}$. From Eq.  (\ref{alpha1}), we   find that $\frac{d\alpha}{dn}=0$ yields  $n=\frac{1}{2}$. Hence, maximum value of $\alpha$ is obtained in the radiation-dominated Universe, as illustrated in Fig. \ref{fig7} (a). Also, in Fig. \ref{fig7} (b), we can see that $\alpha$ increases  with increasing $R_{1}$.

\begin{figure}
 \begin{center}$
 \begin{array}{cccc}
\includegraphics[width=65 mm]{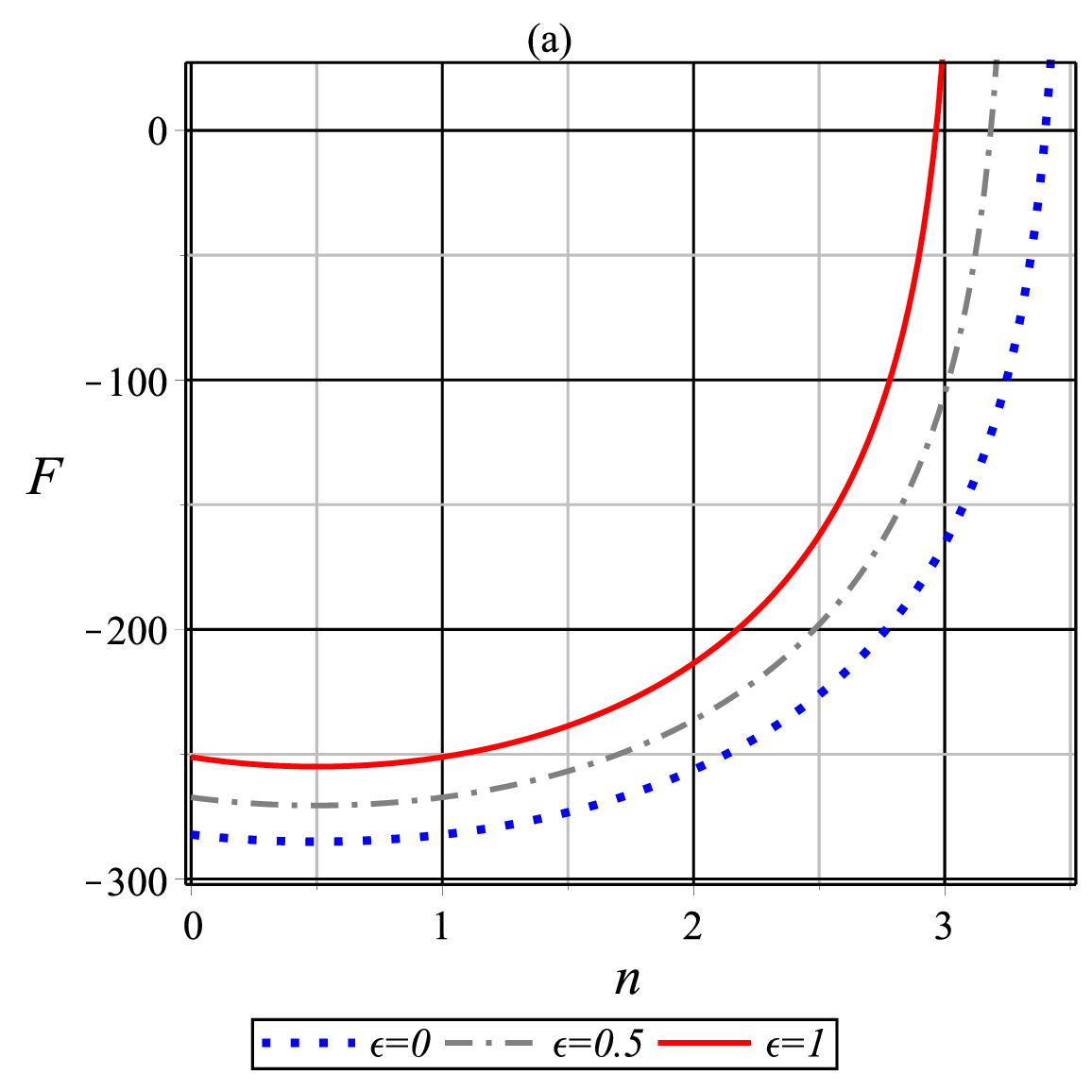}\\
\includegraphics[width=65 mm]{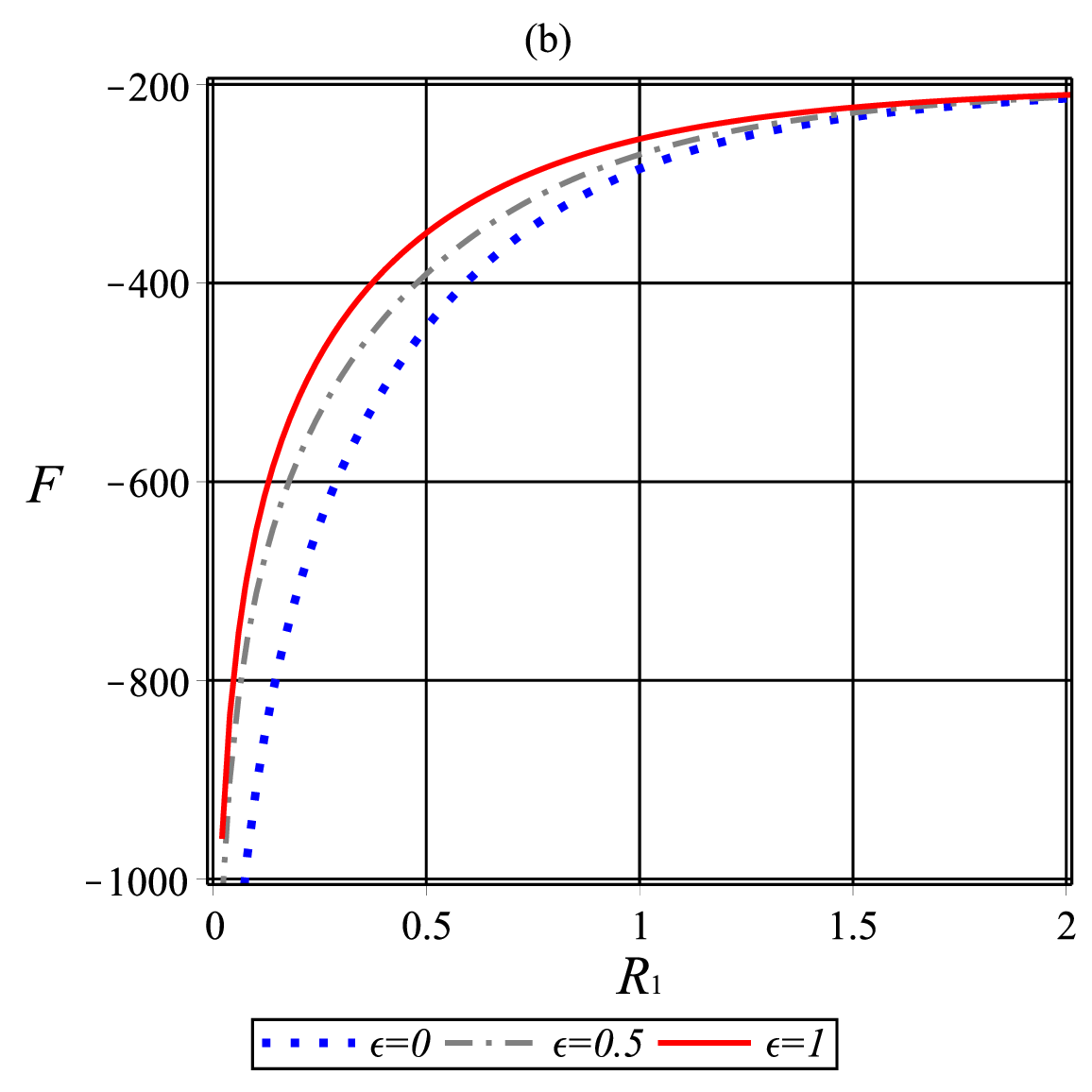}
 \end{array}$
 \end{center}
\caption{Behavior of $F$ for $m=G=\Lambda=1$. (a) In terms of $n$ for $R_{1}=1$;  (b) In terms of $R_{1}$ for $n=0.5$.}
 \label{fig8}
\end{figure}

The maximum of $\alpha$ corresponds to the minimum of the Helmholtz free energy. This  behavior can be obtained from Eq. (\ref{F0}) and  is illustrated in Fig. \ref{fig8}. In fact, we observe   from Fig. \ref{fig8} (a), that the minimum of the Helmholtz free energy, which is equilibrium state of a given system, corresponds to $n=\frac{1}{2}$. Also it is obvious from Fig. \ref{fig8} (b) that the Helmholtz free energy increases by increasing $R_{1}$. Moreover, we can see that for $n<3$ the Helmholtz free energy is negative.

\section{Virial Expansion}
It is possible to study virial expansion for this system of galaxies interacting through a gravitational potential in an expanding Universe (\citealt{ahm02}; \citealt{ahm06}). This virial expansion can be used to obtain the equation of state for this system. Now we will use the gravitational partition function modified by the cosmological constant term to analyze the effect of cosmological constant term on the  equation of state.   Thus, for the case of large galaxy clustering (in the limit $V\rightarrow\infty$), with fugacity $z=\exp ({{\mu}/{T}})$  ($\mu$ as the chemical potential of the given system), one can write
\begin{eqnarray}\label{001}
\frac{P}{T}=\frac{1}{\Lambda_{1}^{3}}\sum_{\nu=1}^{\infty}{I_{\nu}z^{\nu}}, && \frac{N}{V}=\frac{1}{\Lambda_{1}^{3}}\sum_{\nu=1}^{\infty}{\nu I_{\nu}z^{\nu}}.
\end{eqnarray}
It may be noted that here $I_\nu$ is the clustering integral, which is a dimensionless parameter, and is given by
\begin{eqnarray}\label{I}
I_{\nu}=\frac{1}{\nu!\Lambda_{1}^{3(\nu-1)}V}\int{\left(\sum_{ij\neq kl}^{\nu} f_{ij}f_{kl}+\prod_{i\neq j}^{\nu} f_{ij}\right)d^{3}r_{1}\cdots d^{3}r_{\nu}}.
\end{eqnarray}
It is easy to find that $I_{1}=1$.
This clustering integral can be written in terms of   the Mayer function, and it has been observed that such Mayer function    depends on the cosmological constant. So the value of this clustering integral would also depend on the cosmological constant. This can be seen, in the simplest case of $\nu=2$, as we can  write
\begin{eqnarray}\label{I2}
I_{2}&=&\frac{m}{2\Lambda_{1}^{3}}\int_{0}^{R_{1}}{\left(\frac{Gm}{T(r^2+\epsilon^2)^{1/2}}+\frac{\Lambda r^2}{6T}-\frac{\ddot{a}r^2}{2aT}\right)r^{2}dr}\nonumber\\
&=&\frac{Gm^{2}R_{1}}{4T\Lambda_{1}^{3}}\sqrt{R_{1}^2+\epsilon^2}\nonumber\\
&-&\frac{Gm^{2}\epsilon^2}{4T\Lambda_{1}^{3}}\left(\ln{(R_{1}+\sqrt{R_{1}^2+\epsilon^2})}- \ln{\epsilon}\right)\nonumber\\
&+&\frac{m}{4T\Lambda_{1}^{3}}\left(\frac{R_{1}^{5}}{5}\left(\frac{\Lambda}{3}-\frac{\ddot{a}}{a}\right)\right)
\end{eqnarray}
Now we can write the case of $\nu=3$  as follows
\begin{equation}\label{I3}
I_{3}=\frac{1}{6\Lambda_{1}^{6}V}\int{f_{123}\hspace{1mm}d^{3}r_{1}\hspace{1mm}d^{3}r_{2}\hspace{1mm}d^{3}r_{3}},
\end{equation}
where  $f_{123}$  can be expressed as
$ f_{123}\equiv f_{12}f_{13}+f_{12}f_{23}+f_{13}f_{23}+f_{12}f_{13}f_{23}$.
Here we observe that $f_{123}$ can be expressed using $f_{ij}$, and the dependence of $f_{ij}$ on the cosmological constant is known. Thus, we can see that $f_{123}$ explicitly depends on the value of the cosmological constant. We can repeat this procedure for higher values of $\nu$.

\begin{figure}
 \begin{center}$
 \begin{array}{cccc}
\includegraphics[width=40 mm]{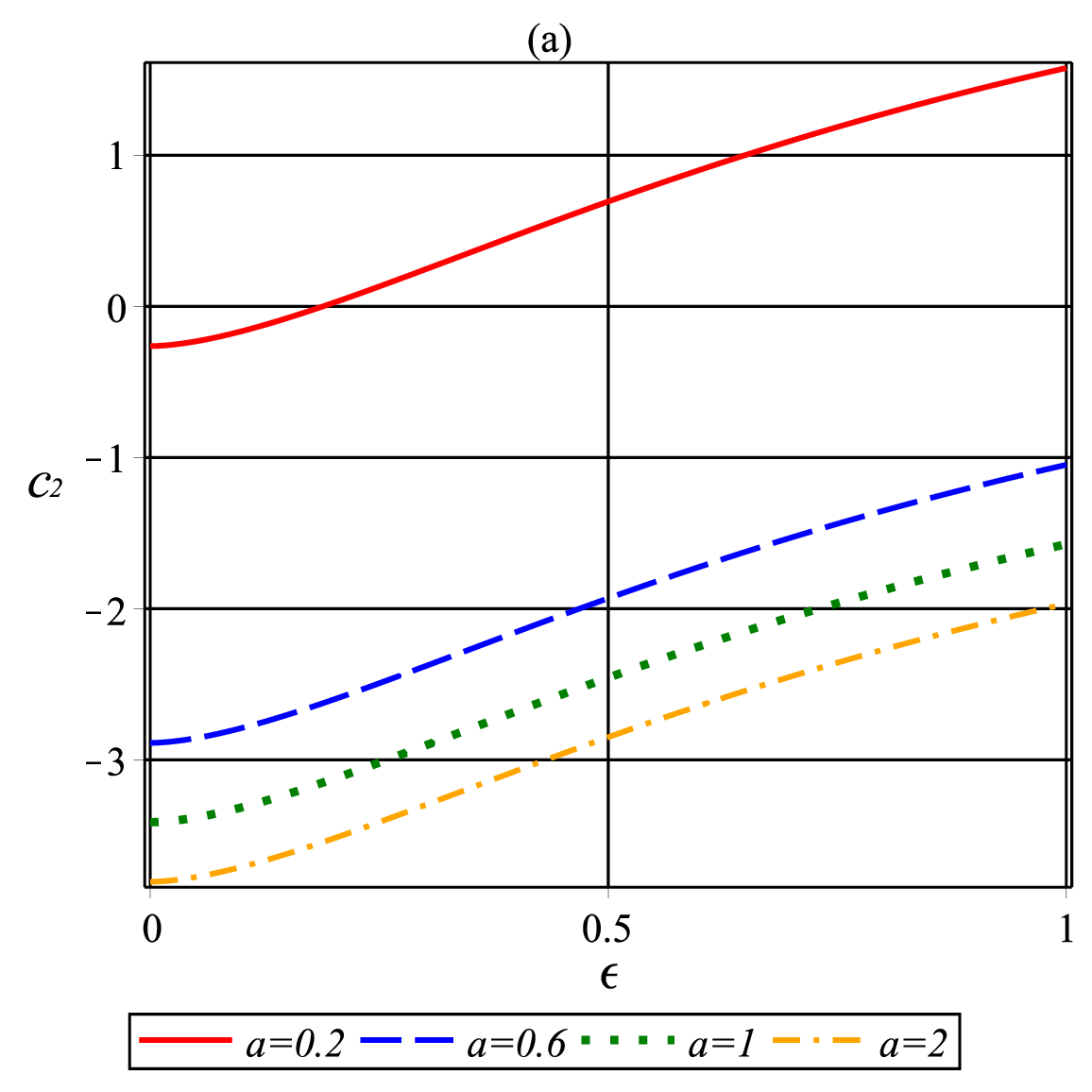}\includegraphics[width=40 mm]{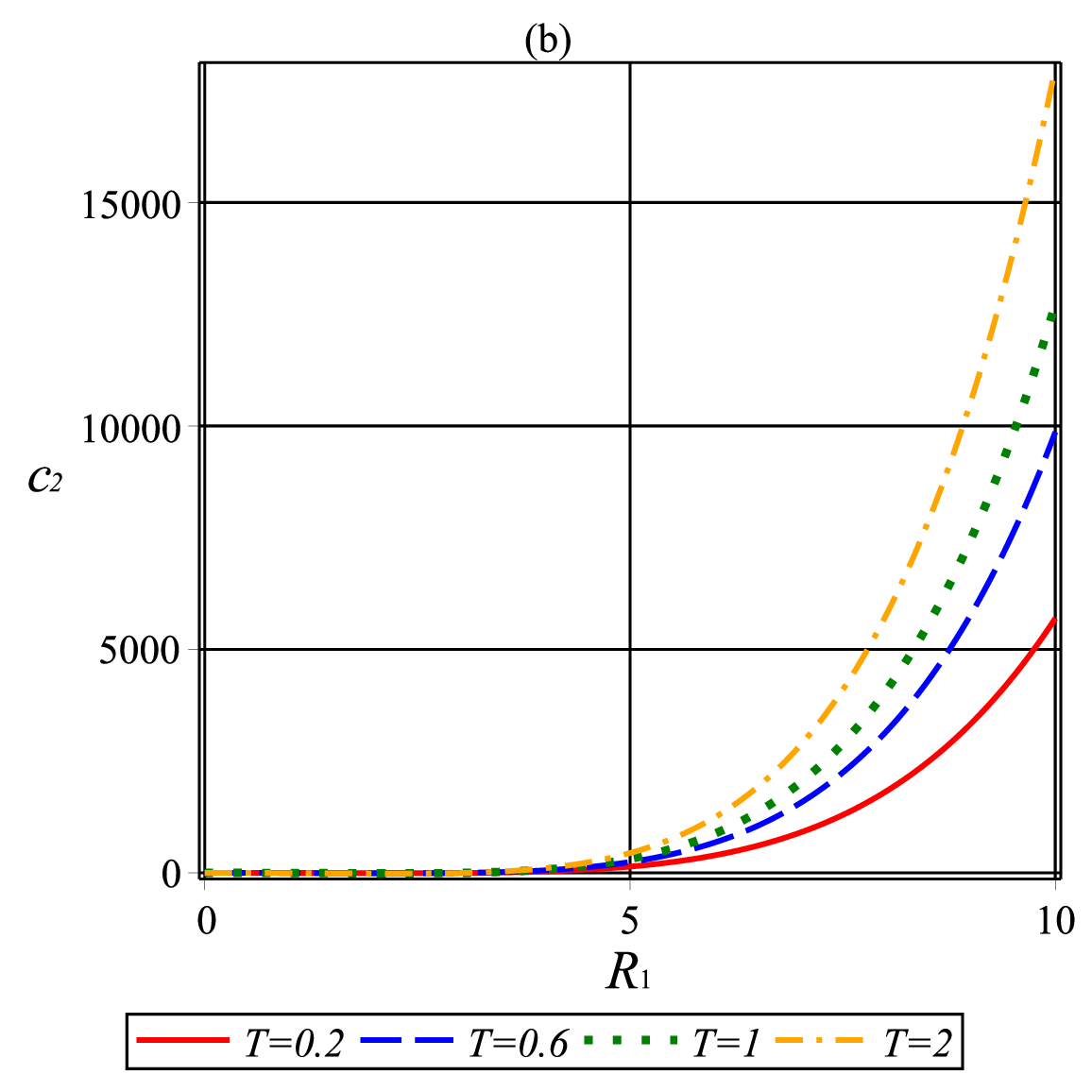}\\
\includegraphics[width=40 mm]{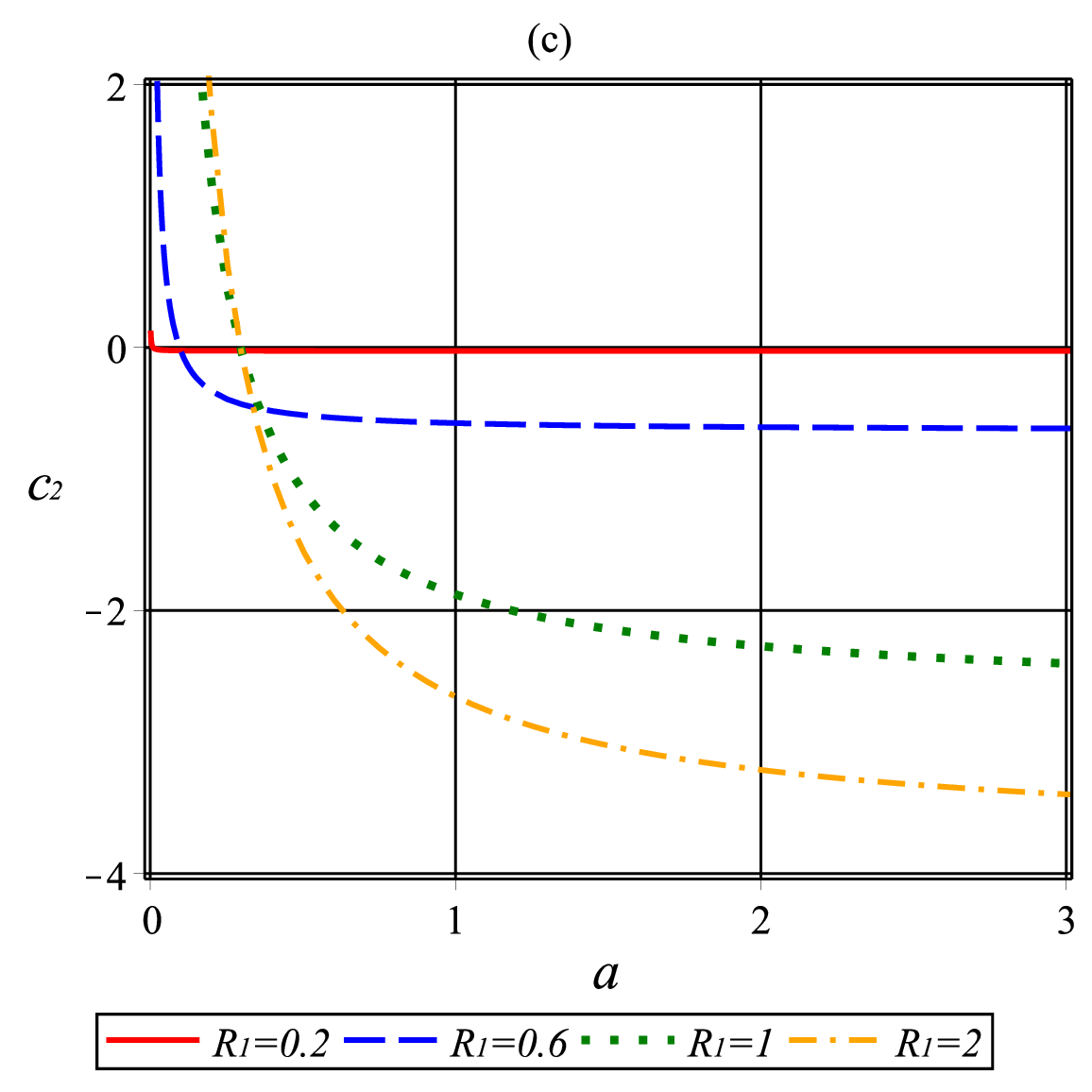}\includegraphics[width=40 mm]{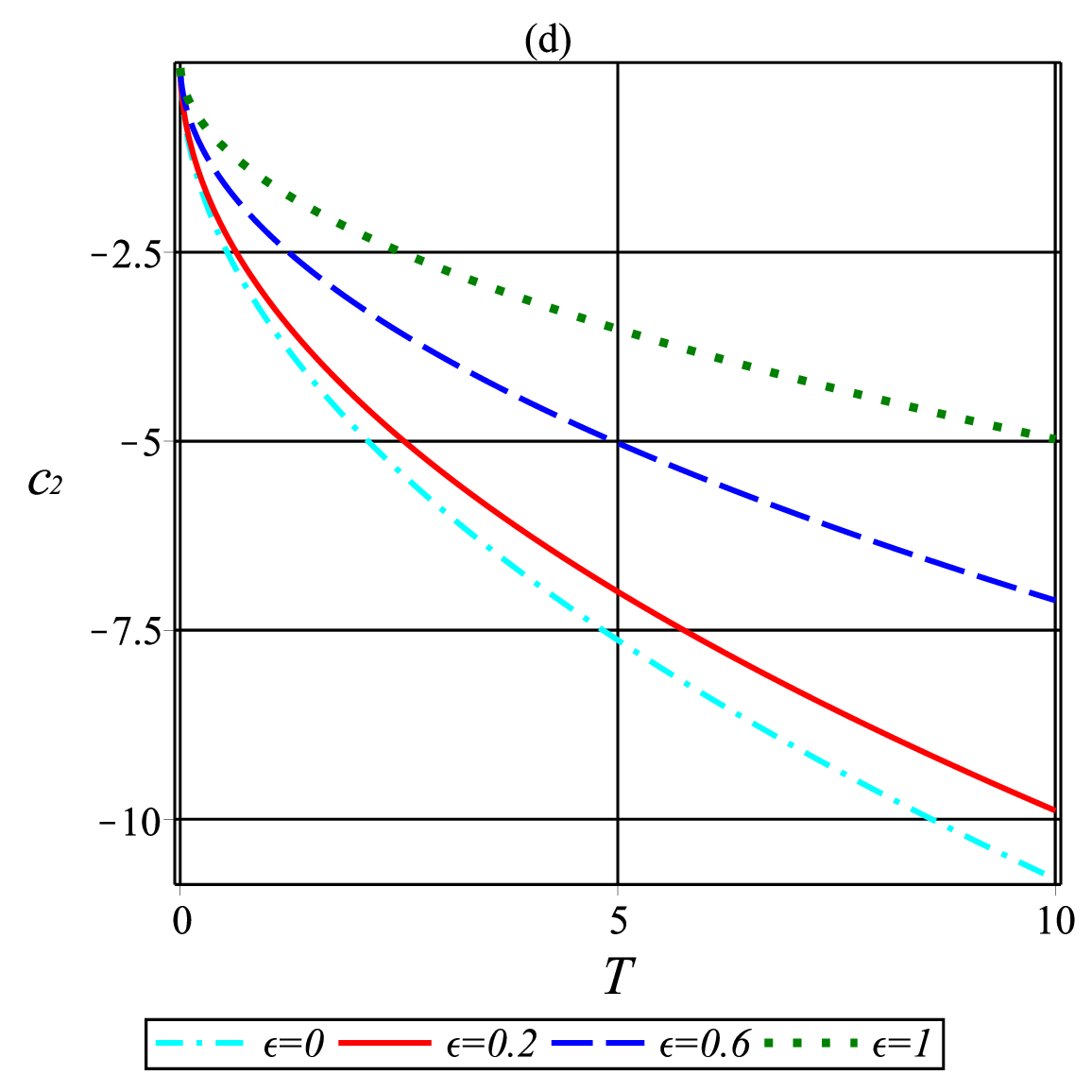}
 \end{array}$
 \end{center}
\caption{Behavior of the second virial coefficient for $m=G=\Lambda=\ddot{a}=1$. (a) In terms of $\epsilon$ for $T=R_{1}=1$;  (b) In terms of $R_{1}$ for $\epsilon=0.5$ and $a=2$; (c) In terms of the $a$ for $T=1$ and $\epsilon=0.8$; (d) In terms of temperature for $a=R_{1}=1$.}
 \label{fig2}
\end{figure}

Eliminating  fugacity $z$ from Eq.  (\ref{001}), one can obtain the clustering equation of state. It has the  following virial expansion
\begin{eqnarray}\label{003}
\frac{PV}{NT}=\sum_{\nu=1}^{\infty}{c_{\nu}(T)\left(\frac{\Lambda_{1}^{3}N}{V}\right)^{\nu-1}},
\end{eqnarray}
where $c_{\nu}(T)$ is called the virial coefficient. In case of $\nu=1$, we obtain first virial coefficient $c_{1}=I_{1}=1$, hence the Eq. (\ref{003}) reduces to the  equation of state for an ideal gas. Other virial coefficients  can also be expressed in terms of the clustering integral. For example, one can obtain
\begin{eqnarray}\label{Ic}
c_{2}&=&-I_{2},\nonumber\\
c_{3}&=&4I_{2}^{2}-2I_{3},\nonumber\\
c_{4}&=&-20I_{2}^{3}+18I_{2}I_{3}-3I_{4}.
\end{eqnarray}
In Fig. \ref{fig2}, we see the behavior of the second virial coefficient $c_{2}$ for the model parameters. In Fig. \ref{fig2} (a), the effect of the softening parameter $\epsilon$ on different values of the scale factor $a$ is shown. It is observed that the $c_{2}$ is an increasing function of $\epsilon$ with both positive and negative values (depending upon scale factor $a$). Now for a small value of  $\epsilon$, we find that parameter $c_{2}$ is approximately a constant. Fig. \ref{fig2} (b) depicts that $c_{2}$ increases as the separation between galaxies increases. In this case, we find that infinitesimal value of  $R_{1}$ produces  negative value for second virial coefficient. Fig. \ref{fig2} (c) shows variation of  $c_{2}$ in terms of scale factor $a$. We observe that it is a decreasing function of $a$,  which approaches  a constant value  for the larger $a$ (late time behavior). Finally, looking at Fig. \ref{fig2} (d), we see that $c_{2}$ is a decreasing function of temperature $T$. It is observed that low $T$ behavior is similar for various values of $\epsilon$.

Now using  the first order approximation (reasonably for $V\rightarrow\infty$), we obtain
\begin{equation}\label{004}
\frac{PV}{NT}=1-\left(\frac{\Lambda_{1}^{3}N}{V}\right)I_{2}+\mathcal{O}(\frac{1}{V^{2}}),
\end{equation}
where $I_{2}$ is given by Eq.  (\ref{I2}). We thus rewrite Eq.  (\ref{I2}) in the following form,
\begin{equation}\label{I2-new}
I_{2}=\frac{f(R_{1})}{2\Lambda_{1}^{3}T},
\end{equation}
where we can write $f(R_{1})$ as
\begin{eqnarray}\label{R2}
f(R_{1})&=&\frac{Gm^{2}}{2}R_{1}\sqrt{R_{1}^2+\epsilon^2}\nonumber\\
&-&\frac{Gm^{2}\epsilon^2}{2}\left(\ln{(R_{1}+\sqrt{R_{1}^2+\epsilon^2})}-\ln{\epsilon}\right)\nonumber\\
&+&\frac{m}{2}\left(\frac{R_{1}^{5}}{5}\left(\frac{\Lambda}{3}-\frac{\ddot{a}}{a}\right)\right).
\end{eqnarray}
This  clustering integral depends explicitly on the cosmological constant and the softening parameter. It may be noted that the softening parameter was introduced to incorporate the extended structure of galaxies in the gravitational partition function (\citealt{ahm10}; \citealt{sas84}; \citealt{ahm02}; \citealt{10}; \citealt{20}; \citealt{ahm06}). This modification of the gravitational partition function in turn modified the value of the  clustering integral. We will now observe that this modification of the cluster modifies the equation of state for this system.
So, with  this new definition,  we write the equation of state (\ref{004}) as
\begin{eqnarray}\label{00500}
\left(P+\frac{a_{v}}{v^{2}}\right)(v-b_v)=T
\end{eqnarray}
where $v={V}/{N}$ is the volume per number of galaxies, and $a_{v}\equiv{f(R_{1})}/{2}$, which is proportional to the clustering integral.
This plays the role of the interaction strength of galaxies (comparing with the Van der Waals gas equation of state). Hence, we can interpret $a_{v}$ as a measure of the strength of  gravitational interaction between galaxies.  It may be noted that the value of $a_{v}$ depends on the cosmological constant, and  the behavior of the system would change due to the cosmological constant term. Furthermore, in absence of any interactions,  such terms would vanish, and the system of galaxies would be approximated by an ideal gas of galaxies. So, the  equation of state for this gas of galaxies would resemble the equation of state for an ideal gas.   In fact, as  $b_{v}$ depends on  the extended structure  of galaxies, it would also vanish if we neglect the effects due to such extended structure. Thus, using such an  approximation,  we note that   $b_{v}=0$. However, due to interactions, the equation of state should resemble the Van der Waals. Thus, we  observe that Eq. (\ref{00500}) is  the usual Van der Waals equation of state for a system of interacting  galaxies. This can be used  for studying  clustering as a phase transition.

\section{Mean field theory of clustering phase transition}
In the previous section, by using virial expansion, we derived  Van der Waals equation of state for a system of interacting galaxies.  This was done by analyzing such a system of interacting galaxies in an expanding Universe.  Furthermore, we also analyzed  the effect of the cosmological constant and  the  scale factor $a$ on clustering. It was also observed that the thermodynamic behavior of this system indicates that there is a gravitational phase transition in it. Now we will analyze such a gravitational phase transition using mean field theory.  Such a   mean field theory for  phase transition in this system  can be analyzed (\citealt{Huang}), using  the Van der Waals equation.
It has been argued that the gravitational   clustering of  galaxies can be regarded as a form of phase transition (\citealt{sin12}; \citealt{sin14}).
The clustering of galaxies occurs by passing through a  mixed phase regime,  in which some parts of the system have clustered, and others have not clustered. So, we can analyze this system as a first order phase transition.
To analyze the clustering of galaxies,  we extend the previous analysis  to a non-zero $b_{v}$, and use the mean field theory to analyze the  phase transition. We note that it is possible to find Landau free energy that can produce  the required behavior,  when it is  minimized  with respect to the order parameter. Here we choose $v$ as the order parameter and $P$ as the conjugate field. Then we can  minimize the  Landau free energy  $\psi(v,P,T)$. Thus, we can write
\begin{eqnarray}\label{dpsi}
\frac{\partial \psi}{\partial v} = 0.
\end{eqnarray}
We can also use
$
\psi(v,P,T)= Pv - v^{-1}{a_{v}}  - T \ln(v-b_{v}).
$ Now minimizing this expression of $\psi(v,P,T)$, we can write
\begin{equation}\label{0051}
P= \frac{T}{(v-b_{v})}  -  \frac{a_{v}}{v^{2}}.
\end{equation}
This is the behavior of the pressure that we had obtained from the Van der Waals equation of state. Now as the average kinetic energy of galaxies would remain the same, the temperature of this system would not change. Thus, we need to analyze the change in the volume for such a system which  can be measured using the isothermal compressibility. Thus, we can write the  isothermal compressibility  of this system as
\begin{equation}\label{iso}
\beta_{T} =- \frac{1}{v}\left(\frac{\partial v}{\partial P}\right)_{T} = \frac{{(1-b_{v})}^2}{v\left(T-2a_{v}(1-b_{v})^2\right)}.
\end{equation}
Approaching the critical point from $T>1$, along  $v=1$, we write the compressibility as
\begin{eqnarray}
\beta_{T} =  \frac{{(1-b_{v})}^2}{\left(\tau +1-2a_{v}(1-b_{v})^2\right)}.
\end{eqnarray}
Here we have  denoted the critical temperature by $T_{c}$ and introduced a dimensionless quantity $\tau$,  such that
\begin{eqnarray}\label{tau}
\tau= \frac{T-T_{c}}{T_{c}}.
\end{eqnarray}
Let us analyze the compressibility in the limit $\tau \rightarrow 0$. We observe that   this gives a finite $\beta_{T}$
\begin{eqnarray}
\beta_{T} =  \frac{{(1-b_{v})}^2}{1-2a_{v}(1-b_{v})^2}.
\end{eqnarray}
Thus, in the limit $\tau \rightarrow 0$, this thermodynamic quantity should have a singular part in addition to the regular part. For the phase transition to take place as $\tau \rightarrow 0$, $\beta_{T}$ should be infinite. So, the    phase transitions can  take place at
$
\tau=2a_{v}{(1-b_{v})}^{2} - 1.
$
Thus, we can write
\begin{eqnarray}
{T_{c}}^{\prime} = T_{c}\left(2a_{v} (1-b_{v})^{2}\right)
\end{eqnarray}
So, in our case, critical temperature changes from $T_{c}$ to ${T_{c}}^{\prime}$. It may be noted that this critical temperature depends on $b_v$, and hence on the extended structure of galaxies. The critical temperature represents the phase transition from a homogeneous phase to a  clustered phase.

Corresponding to a mean field theory in the region of a first order phase transition, Landau free energy $\psi$ must have two minima at volume $v_{1}$ and $v_{2}$. The two conditions (continuity of functions and their derivatives) have to be satisfied as
\begin{eqnarray}
\left(\frac{\partial \psi}{\partial v}\right)_{v=v_{1}}&=&\left(\frac{\partial \psi}{\partial v}\right)_{v=v_{2}},\\
\psi(v_{1}) &=& \psi(v_{2}).
\end{eqnarray}
This  leads to the following equation for temperature
\begin{eqnarray}
T\left(\frac{1}{v_{1} -e} - \frac{1}{v_{2} -e}\right) - a_{v}\left( \frac{1}{{v_{1}}^{2}} - \frac{1}{{v_{2}}^{2}}\right) =0.
\end{eqnarray}
We can also write the following equation for the pressure of this system
\begin{eqnarray}
P(v_{1} -v_{2})= \int_{v_{2}}^{v_{1}} P dv.
\end{eqnarray}

The Maxwell construction is required here as the  volume continues to increase, but the pressure remains constant.
In the  Maxwell construction,   two coexisting volumes $v_{1}$ and $v_{2}$, are symmetrically placed around the critical volume, say $v$.
We can define these two volumes as
$
v_{1}\equiv1+\delta, $ and $ v_{2}\equiv1-\delta,
$, with  $\delta$ as a small parameter.
Hence, by using the Maxwell construction and results obtained in the previous section, we can write
\begin{equation}\label{e}
\frac{2\delta(1+\tau)}{(1-b_{v})^{2}-\delta^{2}}
+a_{v} \left(\frac{1}{{(1+\delta)}^{2}} -\frac{1}{{(1-\delta)}^{2}}\right) =0.
\end{equation}
A possible solution to the Eq. (\ref{e}) can be expressed as
\begin{equation}
\delta\approx\sqrt{\frac{(1+\tau)-2a_{v}{(1-b_{v})}^{2}}{2(1+\tau-a_{v})}},
\end{equation}
where we neglected the terms of $\mathcal{O}(\delta^{4})$.
There is a coexisting volume at phase transition when $\delta \to 0$ corresponding to
$
\tau= 2a_{v}{(1-b_{v})}^{2}-1.
$
Thus, Maxwell construction provides valuable information about the phase transition. Now by neglecting the extended structure of galaxies, we can use $b_{v}=0$, and write $\tau$ as
\begin{eqnarray}
\tau= 2a_{v}-1=f(R_{1})-1=2\Lambda_{1}^{3}T I_{2}-1.
\end{eqnarray}
Hence, under assumption $T=m=1$, we can express $\tau$ as $
\tau= ({I_{2}}/{\sqrt{2\pi^{3}}}) - 1.
$
Therefore, we can observe that $\tau$ is proportional to $T_2$
\begin{eqnarray}
\tau\propto I_{2}.
\end{eqnarray}
In that case, the behavior of $\tau$ is represented by the dotted green line of Fig. \ref{fig2} (b). Thus,   $\tau\rightarrow0$ as  $R_{1}\rightarrow0$, as pointed out before.  It may be noted that it was known that the clustering can be analyzed as a gravitational phase transition in absence of the modification of the gravitational partition function by a  cosmological   constant term (\citealt{sin12}; \citealt{sin14}). However, in this paper, we have demonstrated that it can still be viewed as a phase transition, even after the gravitational partition function was modified by the cosmological constant term. Furthermore, in this paper, this was done using the Maxwell construction.

\section{Cosmic energy equation}
In this section, we discuss cosmic energy equation, as it  can provide important information about clustering of galaxies  (\citealt{clus}; \citealt{PEEBLES}; \citealt{PEEBLESb}).  The cosmic energy equation  has been used  to analyze the effects of cosmic expansion on a large ensemble  of pressure-less galaxies  interacting via a Newtonian gravitational potential
 (\citealt{clus}; \citealt{PEEBLES}; \citealt{PEEBLESb}).   The extended structure of galaxies has been incorporated into the cosmic energy equation using the softening parameter   (\citealt{ce}), and this has been done for other non-point like masses (\citealt{Ahmad}).  A large distance modification to the  Newtonian potential has also been used to study a modification of the cosmic energy equation   (\citealt{Hameeda}).   In the present work, we analyze the modification of the cosmic energy equation by a cosmological constant term.   For a system of galaxies with the internal energy  $U$, pressure $P$, and scale factor $a(t)$, the first law of thermodynamics can be written as
\begin{eqnarray}
\frac{d(U a^3)}{dt}+P\frac{da^3}{dt}=0.
\end{eqnarray}
Writing the equations for energy  $U$ and pressure $P$ in terms of the potential modified by a  cosmological constant term, we obtain
\begin{eqnarray}
U&=&\frac{3}{2}NT+\frac{N\rho}{2}\int_{V}\Phi(r)\xi(r)4\pi r^2dr,\\
P&=&\frac{NT}{V}-\frac{\rho^2}{6}\int_{V}r\frac{d\Phi(r)}{dr}\xi(r)4\pi r^2dr,
\end{eqnarray}
where $\rho$ is density number and $\xi(r)$ is the correlation
function which gives the probability of finding another object in a given radius. The integral for the  correlation
function   over a certain volume is obtained using  the mean
square number fluctuation as
\begin{eqnarray}\label{cor}
\int\xi dV=\frac{(2-b)b}{(1-b)^2},
\end{eqnarray}
where we have used ${\partial b}/{\partial V}=-({x}/{V})({db}/{dx})=-({b(1-b)})/{V}$ and the Eq.  (\ref{x000}). Here $b$ is the clustering parameter, which measures the clustering in the system  (\citealt{ahm02}; \citealt{1b}).
Now by using  Eq.(\ref{potential}), we can write the above parameters as
\begin{eqnarray}
U&=&\frac{3}{2}NT + W_{\epsilon} + W_{M},\\
P&=&\frac{3NT + W_{\epsilon} + {\epsilon}^{2}W_{\epsilon}^{\prime} - 2W_{M}}{3V}.
\end{eqnarray}
Here we can write $W_{\epsilon}, W_{M}, W_{\epsilon}^{\prime}$ as

\begin{eqnarray}
W_{\epsilon}&=&-\frac{GN\rho m^2}{2}\int {\frac{\xi(r)}{(r^2+\epsilon^2)^{\frac{1}{2}}}4\pi r^2dr},\\
W_{M}&=&\frac{N\rho m}{2} {\left(\frac{\ddot a}{2a}-\frac{\Lambda}{6}\right)}\int {r^{2} \xi(r) 4\pi r^2dr},\\
W_{\epsilon}^{\prime}&=&\frac{GN\rho m^2}{2}\int {\frac{\xi(r)}{(r^2+\epsilon^2)^{\frac{3}{2}}}4\pi r^2dr}.
\end{eqnarray}
It may be noted that the contribution from the cosmological constant comes from $W_{M}$, and the original contributions come from $W_{\epsilon}$ and $W_{\epsilon}^{\prime}$  (\citealt{ce}).
Now  it is known that we can write the conservation of energy  for such a system using  the cosmic energy equation (\citealt{Hameeda}; \citealt{Hameeda2}). So, we can write  the conservation of energy  for this system modified by  the cosmological constant  term as
\begin{equation}
\frac{d(K+W_\Lambda)}{dt} + \frac{\dot{a}}{a}\left(2K+ W_\Lambda (1+\eta)\right) = 0,
\end{equation}
where $K$ is the kinetic energy and $W_\Lambda = W_{\epsilon} + W_M$ is the total correlation energy in the presence of cosmological constant $\Lambda$.
This is the cosmic energy equation for $\Lambda$CDM model.

The cosmic energy equation derived above can be  simplified by using the definition of clustering parameter $b_{\Lambda}$, which is the ratio of gravitational correlation energy $W_{\Lambda}$ to kinetic energy $K$,  for a given value of $\Lambda$ (\citealt{ahm10})
\begin{equation}
    b_{\Lambda}=-\frac{W_\Lambda}{2K}.
\end{equation}
It may be noted that we can use $y_{\Lambda}(t)$=${1}/{b_{\Lambda}(t)}$ to simplify the cosmic energy equation (\citealt{sas}; \citealt{sas2}). So, we can write the    cosmic energy equation
for a system  modified by the cosmological constant
as
\begin{eqnarray}\label{e1}
\frac{dy_{\Lambda}(t)}{dt} - \frac{2-y_{\Lambda}}{W_\Lambda}\frac{dW_\Lambda}{dt} - \frac{2\dot{a}}{a}(1-y_{\Lambda} + \eta)=0.
\end{eqnarray}
We use the power law form of the correlation energy
$
W_\Lambda(t) \propto t^{\omega},
$ where  $\omega$ is  a real number. The power law form of the scale factor  given in Eq. (\ref{scale}) can be  used to  solve Eq. (\ref{e1}), and we can write such a solution as
\begin{eqnarray} \label{b}
y_{\Lambda}(t) = y_{c} + (y_{{0}} - y_{c}) \left(\frac{a}{a_{0}}\right)^{j},
\end{eqnarray}
where $j = - (({3\omega}/{2}) + n)/n$ and $y_{{0}} = 1/ b_{0}$. Here $y_{c}={1}/{b_{c}}$ is  critical value of $b$ at which  the  system is virialized.
We can also write
$
\eta = {{\epsilon}^{2} W_{\epsilon}^{\prime} -3W_{\Lambda}}/{W_\Lambda}.
$
We can  use it to determine the critical value of the clustering parameter. So, using $\eta$, we can express this critical value as
$y_{c} =  ({2\omega +2n +2n\eta})/({\omega +2n})$.
Now as for  the power law, we have $\omega \sim {1-\bar{n}}/{3}$, so we can write this critical value as
\begin{eqnarray}\label{73}
b_{c} = \frac {5-\bar{n}}{6-2\bar{n}+4\eta}.
\end{eqnarray}
Thus, it is possible to obtain an explicit expression for $b_c$ in terms of $\bar{n}$ and $\eta$.
However, as  $\eta$ is very small, it  can be ignored.  So, the extended structure   of galaxies does not   contribute to  $b_{c}$. This  indicates that  $b_{c}$ is independent of  such local modifications to the gravitational potential.  It only depends on the value of $\bar{n}$. Now we can obtain   the expression for   $b_{\Lambda}$ from Eq. (\ref{b}) as
\begin{eqnarray}\label{end}
b_{\Lambda}=\frac{b_{c}}{1+\left(\frac{b_{c}}{b_{0}}-1\right)\left(\frac{a}{a_{0}}\right)^{j}}.
\end{eqnarray}

We can use this expression for $b_{\Lambda}$ to analyze the relation between the redshift and clustering.
So, using the relation $1+z = {a_{0}}/{a}$, where $z$ is the red shift and $a_{0}$ is current value of the scale factor, we  can study the variation of $b_{\Lambda}$  with  $z$ for  different models i.e, for different values of $\bar{n}: 1,0,-1,-2$. These different values of $\bar{n}$ correspond to different values of  $\omega$
because $\omega$ is related to $\bar{n}$ as  ${1-\bar{n}}/{3}$. So, for $\bar{n}=2, 1, 0, -1, -2, -3$, the corresponding  values of    $b_{c}$ are $1.5, 1, 0.83, 0.75, 0.7, 0.66$, respectively. Now for  $\bar{n}=3$, $b_{c}$ diverges. It is possible to analyze the  variation of  $b_{\Lambda}(z)$ with $z$, by fixing  $b_{0}$ at a fixed minimum value.

\begin{figure}
 \begin{center}$
 \begin{array}{cccc}
\includegraphics[width=50 mm]{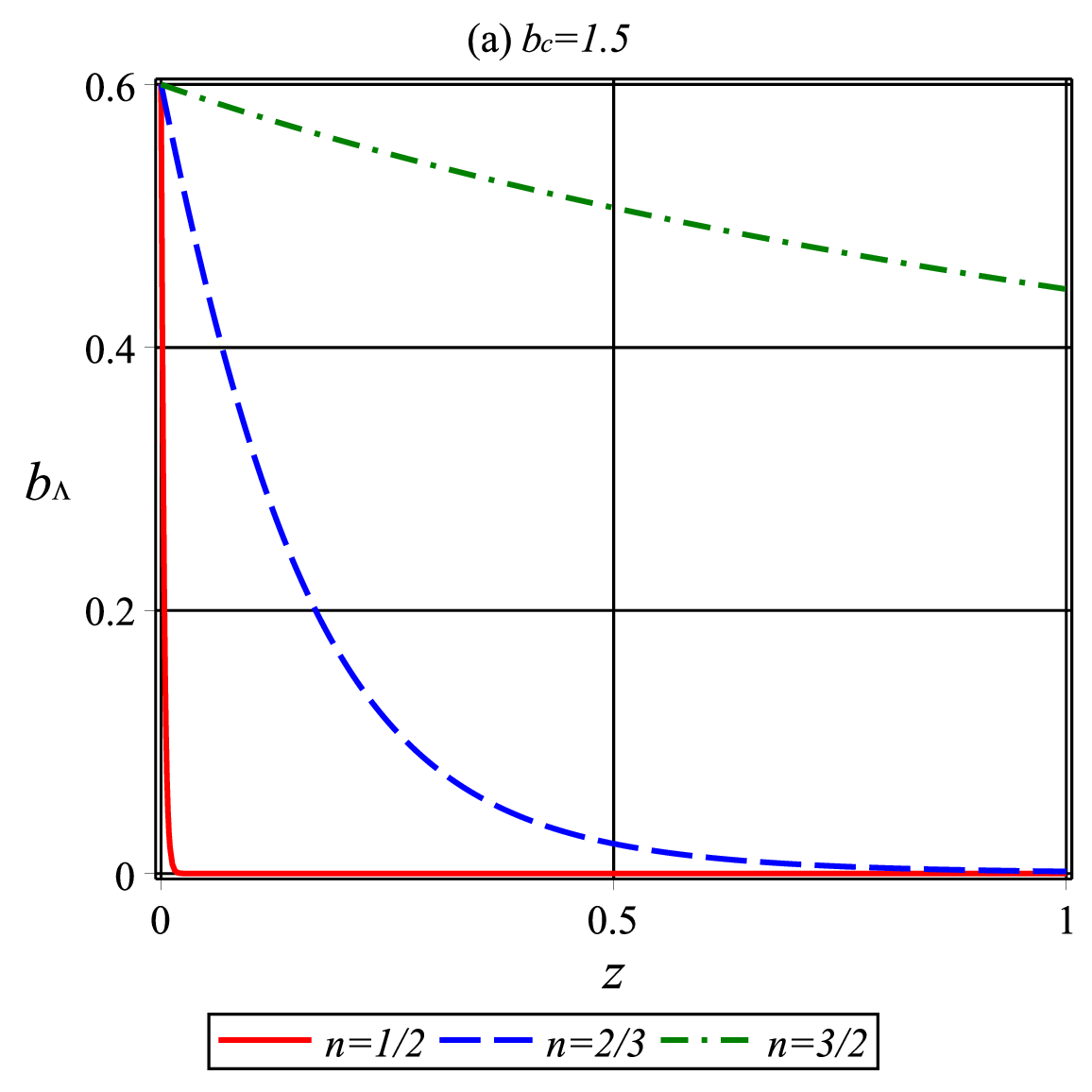}\\
\includegraphics[width=50 mm]{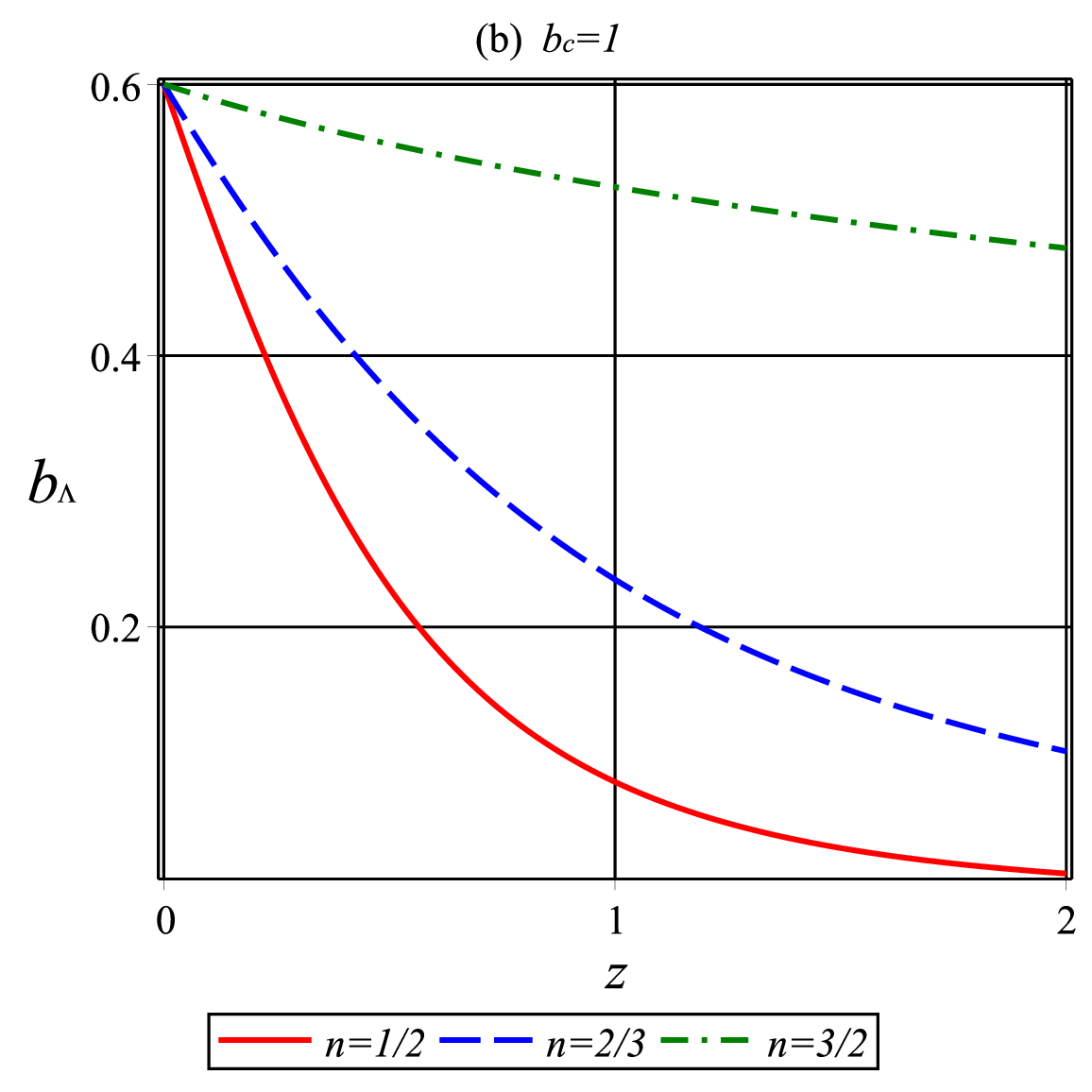}\\
\includegraphics[width=50 mm]{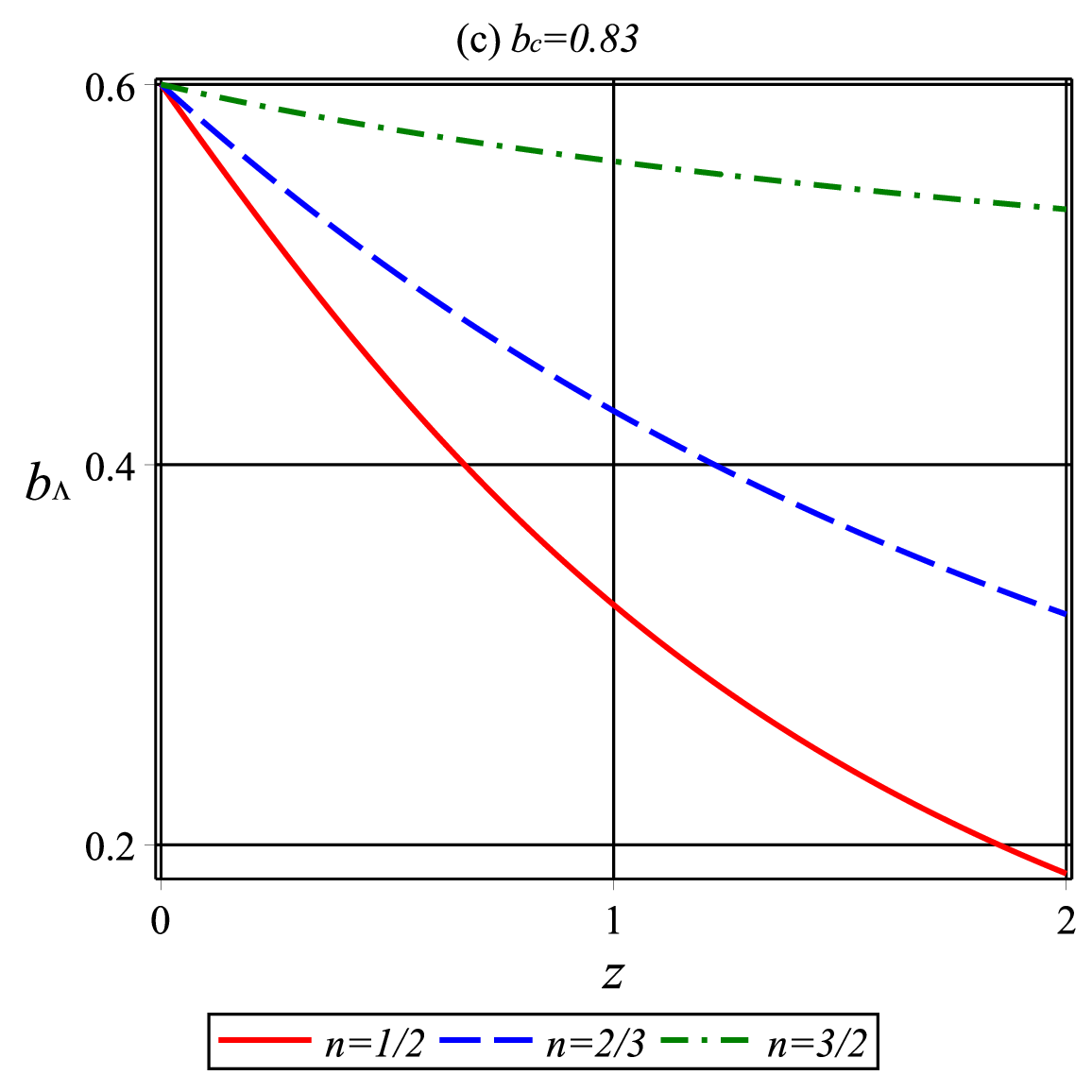}
 \end{array}$
 \end{center}
\caption{Behavior of $b_{\Lambda}$ in terms of the redshift for $b_{0}=0.6$ and $\omega>\omega_{c}$.}
 \label{fig9}
\end{figure}

In Fig. \ref{fig9}, we plot $b_{\Lambda}$ against $z$  to understand its time-dependence. This is done by  fixing $b_{0}=0.6$ (with $G=m=1$).     In  Fig. \ref{fig9} (a),  we analyze the behavior of $b_{\Lambda}$ for  $\bar{n}=2$. Here $\omega=-{1}/{3}$ and so $b_{c}=1.5$.  In a  matter-dominated Universe,  $b_{\Lambda}$ decreases suddenly and  becomes a constant. In Fig. \ref{fig9} (b), we analyze the case  $\bar{n}=1$, with  $\omega=0$, and so $b_{c}=1$. Here  $b_{\Lambda}$ is a decreasing function of $z$. However, it becomes  a constant at high values of the  redshift. In all these three plots, the behavior of $b_{\Lambda}$ corresponding to the radiation-dominated Universe    ($n=1/2$) is  represented by solid red line,  that of a  matter-dominated Universe   is represented by dashed blue line($n=2/3$) and dash dotted green line for  $n=\frac{3}{2}$.
Then, we study the case of $\bar{n}=0$, for which $\omega={1}/{3}$ and so $b_{c}=0.83$. It  is  illustrated in Fig. \ref{fig9} (c).

\begin{figure}
 \begin{center}$
 \begin{array}{cccc}
\includegraphics[width=60 mm]{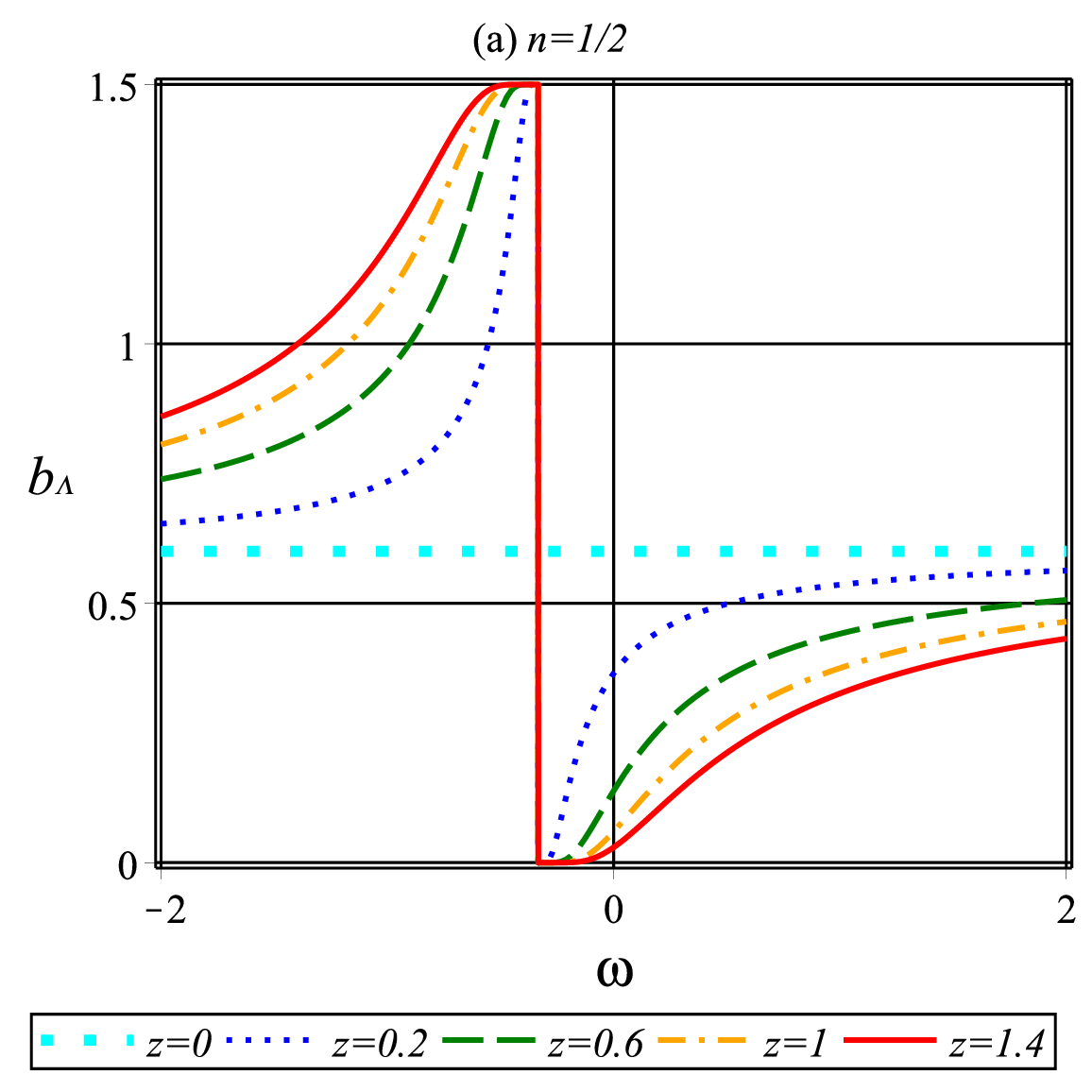}\\
\includegraphics[width=60 mm]{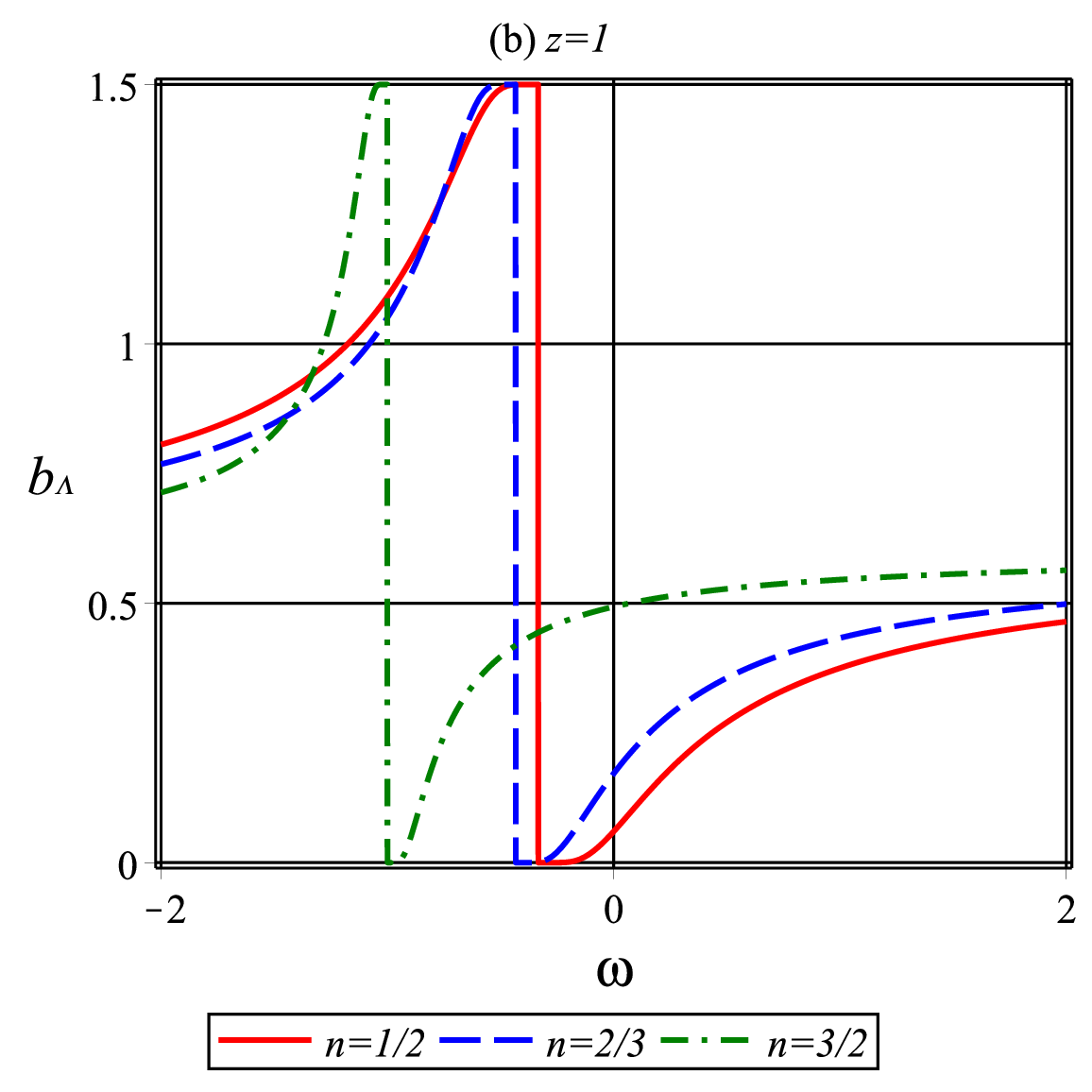}
 \end{array}$
 \end{center}
\caption{Behavior of $b_{\Lambda}$ in terms of the $\omega$ for $b_{0}=0.6$.}
 \label{fig10}
\end{figure}

For all other values mentioned above, we can observe similar behavior i.e., $b_{\Lambda}$ is a decreasing function of redshift for $\omega>\omega_{c}$. We also observe that  $\omega_{c}$ is has a  negative value and  is independent of the redshift ($\omega=\omega_{c}$ is a singular point).  It may be noted that using the   $N$-body simulation (\citealt{MNRAS}), it was observed that  $b_{\Lambda}$ is an increasing function of the redshift. This  is possible  if we chose $\omega<\omega_{c}$. This is illustrated in  Fig. \ref{fig10}, which depicts behavior of $b_{\Lambda}$ in terms of $\omega$. Now   $n={1}/{2}$ produces  $\omega_{c}\approx-0.34$. In Fig. \ref{fig10} (a), we show that $b_{\Lambda}$ is increasing (decreasing) function of redshift for $\omega<\omega_{c}$ ($\omega>\omega_{c}$), and there is a singular point for $\omega=\omega_{c}$.  We see that $b_{\Lambda}$  at $z=0$    is constant. It is  equal to initial value of  $b_{0}=0.6$. The value of $\omega_{c}$ only depends on $n$, which is shown in Fig. \ref{fig10} (b). We can observe that  for the larger values of $n$,  $\omega_{c}$ is smaller.

\begin{figure}
 \begin{center}$
 \begin{array}{cccc}
\includegraphics[width=60 mm]{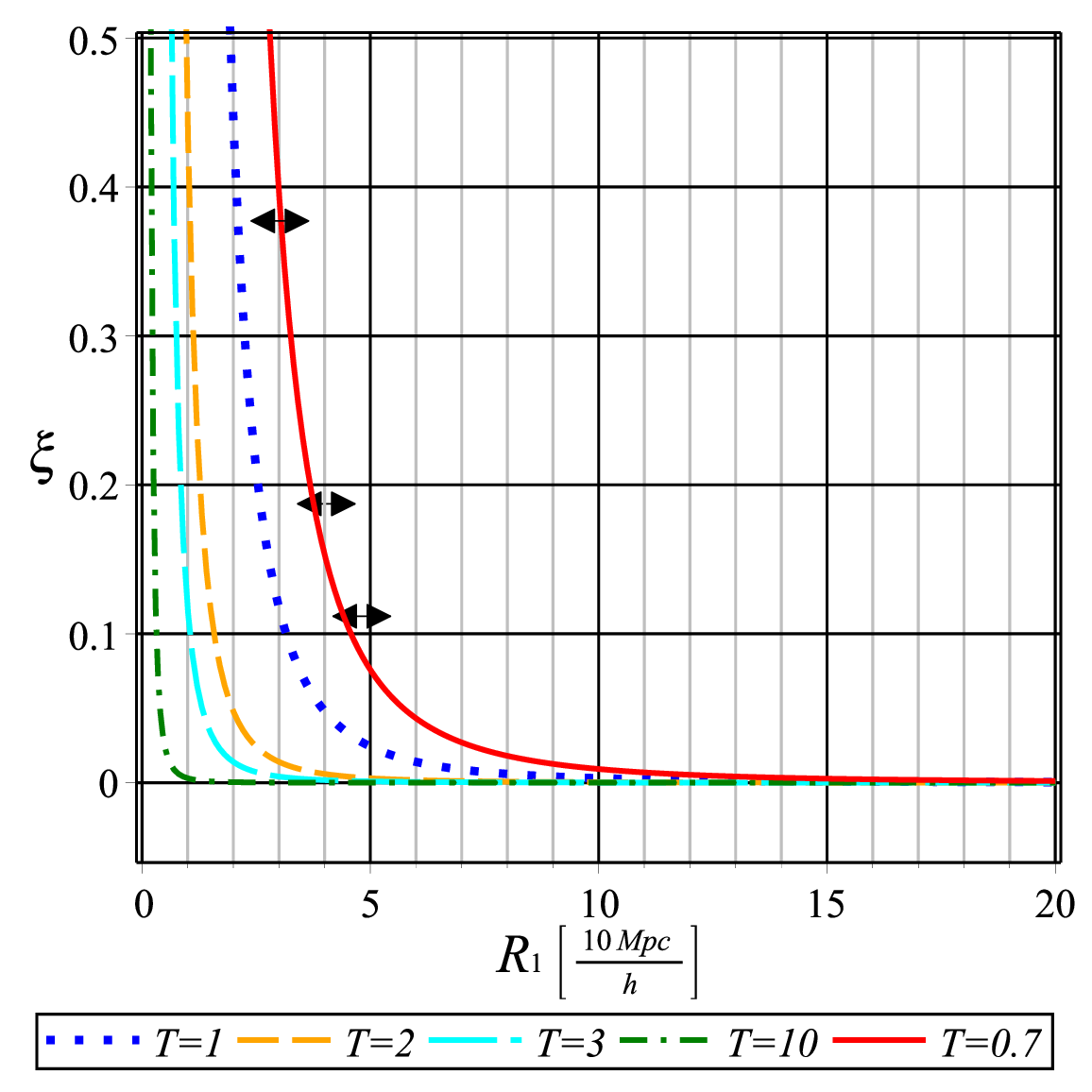}
 \end{array}$
 \end{center}
\caption{Correlation function in unit volume. Two-side arrows show observational data.}
 \label{fig12}
\end{figure}

Now we can compare the results of this paper with observations using  redshifts for (\citealt{Hong}.
The value of $b_\Lambda$  varies from  $0\leq\langle b_{\Lambda}\rangle\leq1$, as expected from physical considerations. However, in  order to check validity of our method,  we can use  some specific values of the  temperature. These values are chosen to numerically analyze the behavior of the system.
We can  use our result from Eqs. (\ref{end}) and  (\ref{73}) to write $b_{c}$ in terms of the temperature.
Then, we obtain a relation for the temperature in terms of the redshift. In order to do that, we use the approximation,  $\eta=AT^{2}$, where $A$ is a constant.
 Now we can plot the  correlation function given by Eq.  (\ref{cor}), and compare it with the observational data (\citealt{Hong}). Two-side arrows of Fig. \ref{fig12}, obtained  from (\citealt{Hong}),  are  consistent with our calculations, for  $T\approx0.7$ (see solid red line of Fig. \ref{fig12}). Thus, it is possible to fix the temperature from observations. It may be noted that this  temperature   is  obtained from a kinetic theory of gases, with each galaxy representing a point like particle of such a gas. This temperature has been fixed from observations of redshifts of galaxies. It has been also shown that the correlation function behaves as a power law on small scales ($R_{1}<5$). Such a behavior has been previously observed for the correlation function (\citealt{Bahcall}).
In fact, we  find the general behavior of the correlation function (see Fig. \ref{fig12}) in agreement with the best-fit $\Lambda$-CDM model (\citealt{Hong}).   It also coincides with $11 103$ and $13 904$ clusters (see Two-side arrows of Fig. \ref{fig12} which are obtained  from \citealt{Hong}), with known redshifts (\citealt{Wen}).

\section{Distribution function}

It is also possible to calculate the distribution function for this system. This distribution function can then be used to analyze the
kinetic energy fluctuations for a system of galaxies. These kinetic energy fluctuations can be used   to relate this model to the observational data.
Now as the galaxies behave as point particles, we can obtain  such distribution functions for this system using the standard methods of statistical mechanics.
Thus, we can write the  probability of finding $N$-galaxies  in the grand canonical ensemble    as
\begin{equation}
F(N)=\frac{\sum_{i}e^{\frac{N\mu}{T}}e^{\frac{-U_{i}}{T}}}{Z_{G}(T,V,z)}=\frac{e^{\frac{N\mu}{T}}Z_{N}(V,T)}{Z_{G}(T,V,z)},
\end{equation}
where $Z_{G}$ is the grand partition function defined by (\citealt{Huang}):
\begin{equation}
Z_{G}(T,V,z)=\sum_{N=0}^{\infty}z^NZ_{N}(V,T)
\end{equation}
and $z=\exp\left({{\mu}/{T}}\right)$ is the activity.

Now this   grand partition function can be expressed as
\begin{equation}
\ln{Z_{G}}=\frac{PV}{T}=\bar{N}(1-b),
\end{equation}
where $\bar{N}$ is the average number of galaxies in a cell. So, we can express  the  probability of finding $N$-galaxies  as
\begin{equation}
F(N)=\frac{\bar{N}(1-b)}{N!}\left(\bar{N}(1-b)+Nb\right)^{N-1}e^{-[\bar{N}(1-b)+Nb]}.
\end{equation}
This is in agreement with the previous works (\citealt{ahm02}; \citealt{sas84}).
The basic assumption for a quasi-equilibrium is that the fluctuations in potential energy over a given volume are proportional to the fluctuations of local kinetic energy. Thus,  for $N$ galaxies and assuming $N$ to be very large, we obtain
\begin{equation}
\frac{Gm^2N(N-1)}{2}\left<\frac{1}{r^{\prime}}\right>=\alpha_1\frac{Nm}{2}v^2,
\end{equation}
where $\left<\frac{1}{r^{\prime}}\right>$ is given by
\begin{eqnarray}
\left<\frac{1}{r^{\prime}}\right>=\left<\left(\frac{1}{(r^2+\epsilon^2)^{\frac{1}{2}}}-\frac{1}{Gm}\left(\frac{\Lambda}{6}-\frac{\ddot a}{2a}\right)r\right)\right>.
\end{eqnarray}
Assuming $\left< N\right>=1$, with $G=m=R=1$, we obtain
\begin{equation}
\alpha_1=\left<\left(\frac{1}{r^{\prime}}\right)\right>\left< v^{2}\right>^{-1},
\end{equation}
where $v$ is the peculiar velocity.

\begin{figure}
 \begin{center}$
 \begin{array}{cccc}
\includegraphics[width=60 mm]{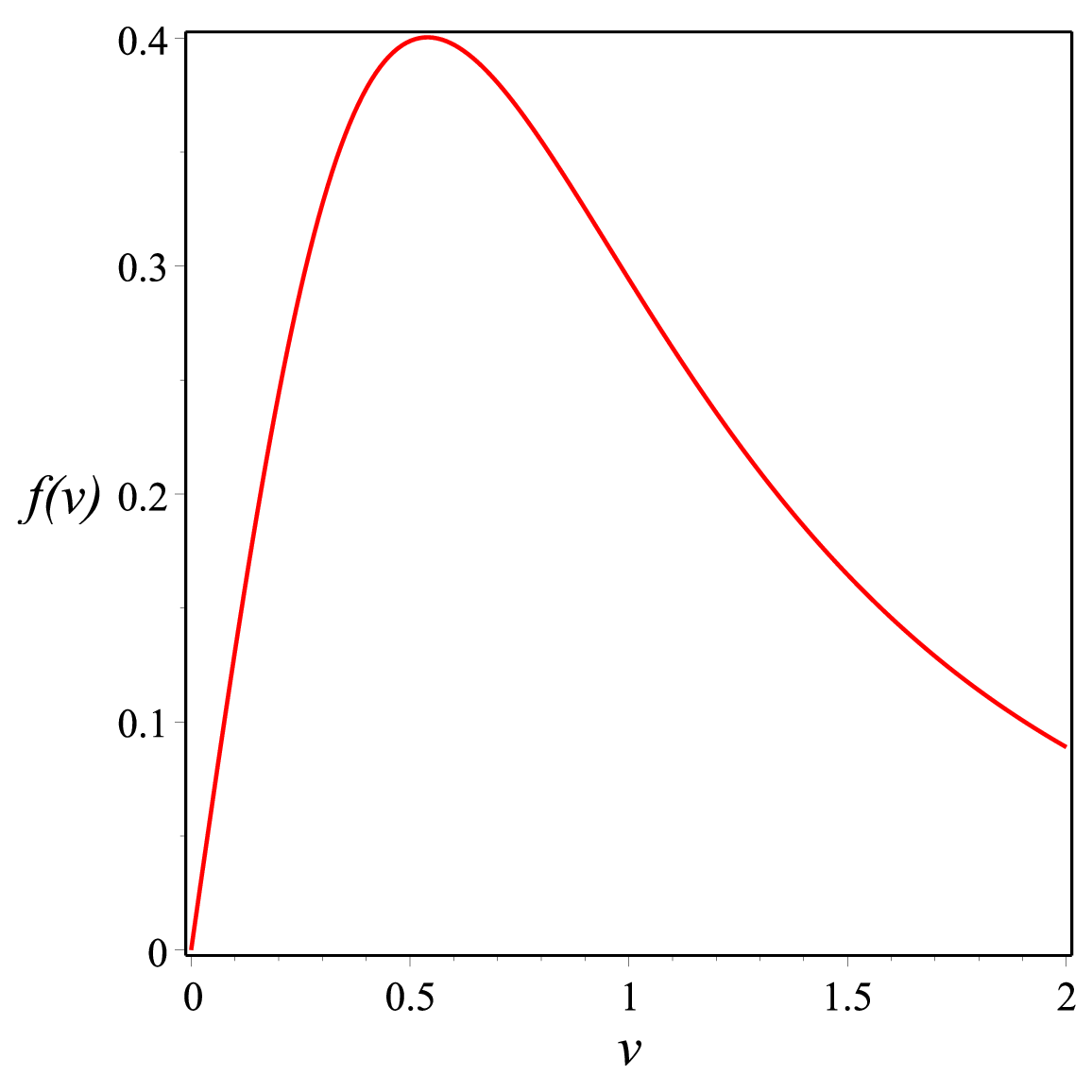}
 \end{array}$
 \end{center}
\caption{Behavior of kinetic energy fluctuations for unit values of the model parameters ($\alpha_{1}=G=m=T=1$).}
 \label{fig13}
\end{figure}

Now, we rescale $F(N)$ from density fluctuations to kinetic energy fluctuations. This is done by replacing $N$ with $N\langle\frac{1}{r^{\prime}}\rangle$ and replacing the average number $\bar{N}$ with $\bar{N}\langle\frac{1}{r^{\prime}}\rangle$. Furthermore, we  substitute $\alpha_1 v^2$ for $N\langle\frac{1}{r^{\prime}}\rangle$ and $\alpha_1\langle v^2\rangle$ for $\bar{N}\langle\frac{1}{r^{\prime}}\rangle$. Finally,  using $N!=\Gamma(N+1)$, we can   express the kinetic energy fluctuations as velocity fluctuations using the Jacobian $2\alpha_1 v$,
\begin{eqnarray}
f(v)&=&\frac{2\alpha_1^2\langle v^2\rangle(1-b)}{\Gamma(\alpha_1v^2+1)}\left[\alpha_1\langle v^2\rangle(1-b)+\alpha_1bv^2\right]^{\alpha_1v^2-1}\nonumber\\
&\times&\exp\left(-\alpha_1\langle v^2\rangle(1-b)-\alpha_1bv^2\right)v.
\end{eqnarray}
 Even though the  modification of the gravitational distribution function from cosmological constant term has been studied (\citealt{cosd91}, here we have explicitly obtained it from the modified gravitational partition function. We also note that the results obtained here are in  agreement with the earlier results  (\citealt{sas2}; \citealt{sasyang}). Thus, a  Gaussian-like distribution (see Fig. \ref{fig13}) is in agreement with earlier  observations (\citealt{Raychaudhury}). Here we have been able to show that this behavior still holds for the system even after the gravitational partition function has been modified by a cosmological constant term. Though this was physically expected, here it has been explicitly demonstrated.

\section{Summary}\label{summary}
In this paper, we have analyzed the effects of expansion of the Universe  on the structure formation in the Universe. This was done by using the gravitational partition function.
As the distance between galaxies is much larger than the size of the galaxies, it is possible to  approximate galaxies as point particles in this  gravitational partition function. These point particles interact through a gravitational potential. The gravitational force pulls these galaxies towards each other, leading to the formation of large scale structure in our Universe. However, the expansion of the Universe   moves these galaxies away from each other. Thus, it is important to analyze the effects of the expansion of the Universe   on the structure formation in our Universe. This can be done by incorporating a cosmological constant term in this gravitational partition function. We have also  used this  gravitational partition function with a cosmological constant term to study the  thermodynamics for this system.  Then, we have used  the virial expansion
to  obtain equation of state for this system. We have  modeled this system of galaxies  as a Van der Waals gas. Here the Van der Waals term was obtained from the interaction of galaxies with each other.  It is interesting to note that Van der Waals behavior in a different context has been studied for clustering of galaxies  (\citealt{Desjacques et al.}).
We have also analyzed a gravitational phase transition in this system. This was done by using  the  mean field theory for this system of galaxies. We have also analyzed the effect of cosmological  constant on the cosmic energy equation, which was later used for   analyzing  the time evolution of the clustering parameter. We have also compared our model with the observational data and used it to constrain the free parameters of our model. This was done using both the cosmic energy equation and distribution function.

It may be noted that the modification of gravitational partition function from $f(R)$ gravity has also been analyzed (\citealt{2}). It was observed that this modified partition function was consistent with the observations. It would be interesting to analyze the gravitational phase transition for this system. The details  of such a  gravitational phase transition would depend on the specific kind of $f(R)$ gravity model. It would also be interesting to obtain the cosmic energy equation for this modified gravitational partition function and use it for analyzing the effects of $f(R)$ gravity on the time evolution of the clustering parameter.  By choosing   different models of $f(R)$ gravity, it is plausible to analyze clustering and the dependence of clustering parameter on those models. Furthermore, the correlation between galaxies can be studied in those models, which can later be compared with observations and then used to constrain the free parameters in $f(R)$ gravity models.
As it is possible to use the gravitational partition function for analyzing clustering in MOND (\citealt{3}), it would be interesting to analyze gravitational phase transition using MOND. It would also be possible  to perform such a calculation for MOG, as MOG  predicts a large scale modification of  gravitational potential (\citealt{4}).  It would be interesting to obtain the distribution of different galaxies for such modified theories of gravity, and then compare it with observations. These observations can then be used to constrain certain free parameters in these models of modified gravity.
It may be noted that it is possible to obtain large scale correction to the gravitational potential using brane-world models and then use this modified gravitational  potential to analyze gravitational partition function for brane-world models (\citealt{Hameeda2}). These large scale corrections to the gravitational potential are  obtained from the super-light brane-world perturbative modes.
This gravitational partition function can then be used to  obtain the dependence of the clustering  on large extra dimensions. These effects could be observed in a system of galaxies,  and so  clustering of galaxies  can be used to constrain the size of such large extra dimensions.

\section*{Acknowledgement}
Authors would like to thank MB Shah for useful discussions.

\end{document}